%% file: main.tex
\documentclass{aastex631}

\newcommand{\codv}[1]{\nabla_{#1}} 		
\newcommand{\paren}[1]{\left(#1\right)}
\newcommand{\colch}[1]{\left[#1\right]}

\usepackage{physics}
\usepackage{graphicx}

\begin{document}

\title{Quadrupole Moment of a Magnetically Confined Mountain on an Accreting Neutron Star in General Relativity}

\author[0000-0003-0034-4615]{Pedro H. B. Rossetto}
\affiliation{School of Physics, University of Melbourne, Parkville, VIC 3010, Australia}
\affiliation{ARC Centre of Excellence for Gravitational Wave Discovery (OzGrav), University of Melbourne, Parkville, VIC 3010, Australia}

\author[0000-0003-2796-5323]{Jörg Frauendiener}
\affiliation{Department of Mathematics \& Statistics, University of Otago, 730 Cumberland Street, Dunedin 9016, New Zealand}

\author[0000-0003-4642-141X]{Andrew Melatos}
\affiliation{School of Physics, University of Melbourne, Parkville, VIC 3010, Australia}
\affiliation{ARC Centre of Excellence for Gravitational Wave Discovery (OzGrav), University of Melbourne, Parkville, VIC 3010, Australia}



\begin{abstract}

\input{Sections/0-abstract}

\end{abstract}



\section{Introduction}
\label{sec:intro}
\input{Sections/1-introduction}

\section{Hydromagnetic equilibrium of a polar mountain}
\label{sec:model}
\input{Sections/2-theory}

\section{Quadrupole moment and ellipticity}
\label{sec:cgw}
\input{Sections/3-cgw}

\section{Discussion}
\label{sec:discussion}
\input{Sections/4-discussion}

\section*{Acknowledgements}
\input{Sections/acknowledgements}


\bibliography{CW_paper.bib}{}
\bibliographystyle{aasjournal}



\end{document}

%% file: Sections/0-abstract.tex
General relativistic corrections are calculated for the quadrupole moment of a magnetically confined mountain on an accreting neutron star.  The hydromagnetic structure of the mountain satisfies the general relativistic Grad-Shafranov equation supplemented by the flux-freezing condition of ideal magnetohydrodynamics, as in previous calculations of the magnetic dipole moment. It is found that the ellipticity and hence the gravitational wave strain are up to $12\%$ greater than in the analogous Newtonian system. The direct contribution of the magnetic field to the nonaxisymmetric component of the stress-energy tensor is shown to be negligible in accreting systems such as low-mass X-ray binaries.

%% file: Sections/1-introduction.tex
Fast-spinning deformed neutron stars are prime candidates for the emission of continuous gravitational waves (CWs) \citep{riles2023Searches}. The Laser Interferometer Gravitational-Wave Observatory (LIGO), Virgo, and Kamioka Gravitational Wave Detector (KAGRA) routinely conduct targeted narrowband searches for CWs from known pulsars in the LIGO Scientific Collaboration data \citep{abbott2022Modelbased, abbott2022Narrowband}, directed searches in supernova remnants \citep{abbott2021Searches} and all-sky searches for unknown CW sources \citep{abbott2022Allsky}. No CW signal has been found so far. It is important to model from first principles the formation mechanisms of non-axisymmetric stellar deformations, such as `mountains', which are static in the star's rotating frame, in order to predict the detectability of the associated CW signal. Proposed mechanisms include elastic stresses in the crust \citep{ushomirsky2000Deformations, gittins2020Modelling, gittins2021Modelling}, single and repeated crustal fracture \citep{giliberti2019Modelling, giliberti2022Starquakes, kerin2022Mountain}, the Lorentz force in the magnetized interior \citep{bonazzola1996Gravitational, haskell2008Modelling} and the Lorentz force in the surface layers of polar mountains confined magnetically on accreting neutron stars \citep{melatos2001Hydromagnetic, pm04, melatos2005Gravitational,rossetto2023Magneticallya}. The latter mechanism is the focus of this paper.

Magnetically confined mountains are formed by the accretion of conducting plasma onto the magnetic polar cap of a neutron star in systems like low-mass X-ray binaries (LMXBs). Several features have been incorporated previously into the theoretical model, including mountain stability \citep{vigelius2008Threedimensional, mukherjee2012Phasedependent, mukherjee2013MHD}, Ohmic and thermal relaxation \citep{vigelius2009Resistive, suvorov2019Relaxation}, core superconductivity \citep{passamonti2014Quasiperiodic,sur2021Impact}, mountain sinking \citep{choudhuri2002Diamagnetic,wette2010Sinking}, various equations of state \citep{priymak2011Quadrupole,mukherjee2017Revisiting}, triaxial configurations \citep{singh2020Asymmetric}, higher magnetic multipole moments and toroidal fields \citep{suvorov2020Recycled, fujisawa2022Magnetically}. Recently, \cite{rossetto2023Magneticallya} formulated the problem of magnetically confined mountains on neutron stars in general relativity. They found that relativistic corrections change the hydromagnetic equilibrium of the accreted matter, diminishing the screening of the magnetic moment three-fold when compared to the Newtonian formulation.

In this paper, we discuss how general relativistic corrections modify the mass quadrupole moment of a magnetically confined mountain and hence affect the emission of gravitational radiation. In Section \ref{sec:model}, we summarize the relativistic model of the system and the numerical method. In Section \ref{sec:cgw}, we calculate the ellipticity of the star, breaking down the contributions from various physical effects. We contrast our findings with the respective Newtonian results. Finally, in Section \ref{sec:discussion}, we estimate the amplitude of the associated gravitational radiation and discuss its detectability.

%% file: Sections/2-theory.tex
We employ the same theoretical framework presented by \cite{rossetto2023Magneticallya}. The system is modelled as an isothermal magnetically confined mountain on a neutron star produced by magnetic burial. We assume that the neutron star generates a background Schwarzschild metric and that the accreted matter and magnetic field adjust to this curved spacetime. Throughout the paper, we assume the fiducial value of $1.4M_\odot$ for the gravitational mass $M_*$ of the neutron star. We assume that the whole system is axisymmetric and, hence, that the magnetic field $B^a$ of the neutron star is fully determined by a scalar function $\psi$ as
\begin{equation}
    \label{mag_field}
    B^a = \frac{1}{r^2\sin^2\theta}\epsilon^{abcd}u_d\codv{b}\phi\codv{c}\psi ,
\end{equation}
where $\epsilon^{abcd}$ is the completely antisymmetric tensor. Under these conditions, the equilibrium of the mountain is described by the relativistic isothermal Grad-Shafranov equation
\begin{equation}
    \label{iso_grad_shaf_eqn}
    \Delta^*\psi=-F'(\psi)\exp[-(1+c_{\rm s}^{-2})(\Phi-\Phi_0)],
\end{equation}
where $c_{\rm s}$ is the speed of sound, $\Phi$ is the relativistic gravitational potential, $\Phi_0$ is a reference potential, $\Delta^*$ is the relativistic Grad-Shafranov operator
\begin{equation}
    \label{rel_gs-operator}
    \Delta^*\psi = \frac{1}{r^2\sin^2\theta}\left\{\pdv{r}\colch{\paren{1-\frac{2M_*}{r}}\pdv{\psi}{r}}
    +\frac{\sin\theta}{r^2}\pdv{\theta}\paren{\frac{1}{\sin\theta}\pdv{\psi}{\theta}}\right\},
\end{equation}
and $F(\psi)$ is a function related to the fluid pressure at the altitude of the reference potential $\Phi_0$.

\cite{pm04} developed a recipe to calculate the function $F(\psi)$ uniquely in the Newtonian theory, by connecting the pre- and post-accretion states through the flux freezing condition of ideal magnetohydrodynamics. \cite{rossetto2023Magneticallya} adapted the recipe to general relativity, obtaining
\begin{equation}
    \label{F_gp_gr}
    F(\psi)=\paren{\dv{M}{\psi}}^{1+c_{\rm s}^2}\paren{\frac{2\pi}{c_{\rm s}^2}\int_C r\sin\theta|\codv{} \psi|^{-1}e^{-\paren{\Phi-\Phi_0}/c_{\rm s}^2}\dd{s}}^{-(1+c_{\rm s}^2)}.
\end{equation}
In \eqref{F_gp_gr}, $\dd{M}/\dd{\psi}$ specifies how accreted mass is distributed along flux tubes labelled uniquely by $\psi$. We use 
the same phenomenological prescription for this function as \cite{pm04,priymak2011Quadrupole}, viz.
\begin{equation}
    \label{M_psi}
    M(\psi)=\frac{M_a}{2}\frac{1-e^{-\psi/\psi_a}}{1-e^{-\psi_*/\psi_a}},
\end{equation}
with $\psi_*=\psi(R_*, 0)$, where $R_*$ is the neutron star's radius and $\psi_a$ is the flux surface that touches the inner edge of the accretion disk. 

\begin{figure}
    \centering
    \includegraphics[width=0.9\textwidth]{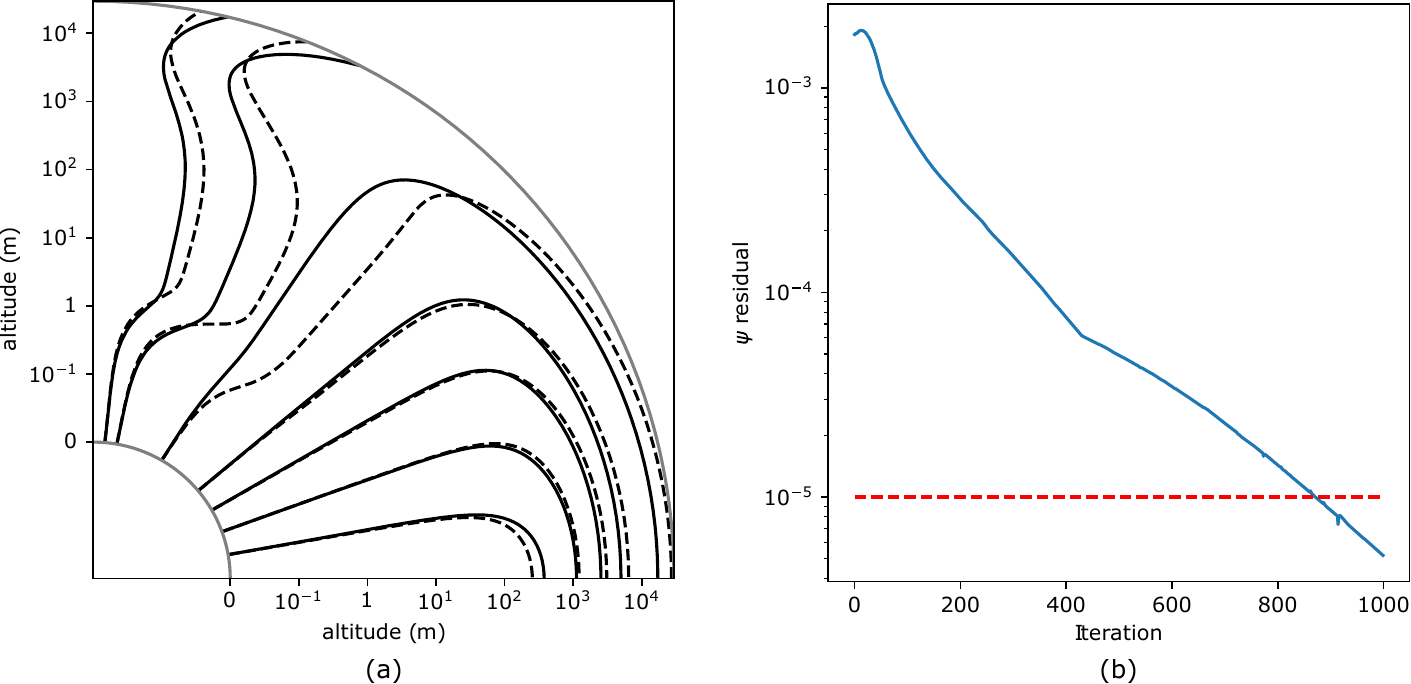}
    \caption{General relativistic versus Newtonian hydromagnetic equilibria for $M_a=10^{-5}M_\odot$ and $b=\psi_*/\psi_a=10$. (a) Magnetic field lines (contours of $\psi$). Solid lines indicate the relativistic solution and dashed lines indicate the Newtonian solution. (b) Residuals of the numerical solver, defined as the average across the grid of the fractional difference between $\psi$ in one iteration and in the next. The dashed red line indicates a tolerance of $10^{-5}$.}
    \label{fig:psi}
\end{figure}

Numerical solutions to equations \eqref{iso_grad_shaf_eqn}--\eqref{M_psi} are obtained iteratively \citep{rossetto2023Magneticallya}. Figure \ref{fig:psi} shows an example of a hydromagnetic equilibrium generated by the solver, for $M_{\rm a}=10^{-5}M_\odot$ and $b=\psi_*/\psi_{\rm a}=10$. Panel \ref{fig:psi}(a) shows the general relativistic equilibrium state for the magnetic field lines ($\psi$ level curves) as solid curves, and the Newtonian equivalent as dashed curves. The system reaches a deformed equilibrium state for both theories, but the deformation in the relativistic case is smaller. Panel \ref{fig:psi}(b) confirms that the numerical $\psi$ residuals decrease with iteration number in a controlled fashion; that is, the numerical solution converges.

%% file: Sections/3-cgw.tex
In the linear regime for the generation of gravitational waves \citep{maggiore2007Gravitational}, the mass quadrupole source term is proportional to the second time derivative of the mass quadrupole moment, defined by
\begin{equation}
    \label{quad_mom}
    Q_{ij} = \int_{V'}T^{00}\paren{{x'}_i{x'}_j-\frac{1}{3}{x'}^k{x'}_k\delta_{ij}}\dd[3]x',
\end{equation}
where $x'^i$ is a spatial position vector (assuming a small source), $V'$ is the volume of the source and $\delta_{ij}$ is the identity tensor. Note that equation \eqref{quad_mom} is usually termed the mass quadrupole moment to distinguish it from the current quadrupole moment, which is proportional to the fluid velocity and generates current quadrupole gravitational radiation \cite{thorne1980Multipole}. However, it is important to recognise that $T^{00}$ in equation \eqref{quad_mom} includes contributions from the energy density in the magnetic field as well as the mass density of the fluid. We discuss this in Section \ref{sec:mag_ellip}. $T^{00}$ is given by
\begin{equation}
    \label{T00}
    T^{00} = e + \frac{B^2}{2}.
\end{equation}
where $e$ is the energy density of the fluid, and we have $B^2=B_aB^a$. Furthermore, $e$ can be decomposed as $e=\rho(1+\varepsilon)$ where $\rho$ is the rest-mass density of the fluid and $\varepsilon$ is its specific internal energy.  We keep both terms in \eqref{T00} for the sake of generality and compare their magnitudes below.

In the triaxial model of a rigidly rotating neutron star, the gravitational wave amplitude is given by \citep{maggiore2007Gravitational}
\begin{equation}
    \label{strain_cw}
    h_0 = \frac{16\pi^2 I_{0}f^2\epsilon}{D},
\end{equation}
where $I_0=(2/5)M_*{R_*}^2$ is the moment of inertia of the star before accretion, $f$ is the rotation frequency, $D$ is the distance to the source and $\epsilon$ is the (mass-energy) ellipticity of the star given by

\begin{equation}
    \label{epsilon}
    \epsilon=\frac{\pi}{I_0}\int_{V'} T^{00} r^4\sin\theta(3\cos^2\theta-1)\dd{r}\dd{\theta}.
\end{equation}

\noindent We also define the material and magnetic ellipticities $\epsilon_e$ and $\epsilon_B$ in terms of the same integral as in \eqref{epsilon} but replacing $T^{00}$ with $e$ and $B^2/2$ respectively to give $\epsilon=\epsilon_e+\epsilon_B$. In equilibrium, $e$ and $B^2$ can be related to the flux-function $\psi$ output from the numerical solver, with
\begin{align}
    &e = \frac{F(\psi)}{c_{\rm s}^2}\exp[-(1+c_{\rm s}^{-2})(\Phi-\Phi_0)],
    \label{e_psi}
    \\
    &B^2 = \frac{1}{r^2\sin^2\theta}
    \colch{\frac{1}{r^2}\paren{\pdv{\psi}{\theta}}^2+e^{2\Phi}\paren{\pdv{\psi}{r}}^2}.
    \label{B_mag}
\end{align}

\begin{figure}
    \centering
    \includegraphics[width=0.9\textwidth]{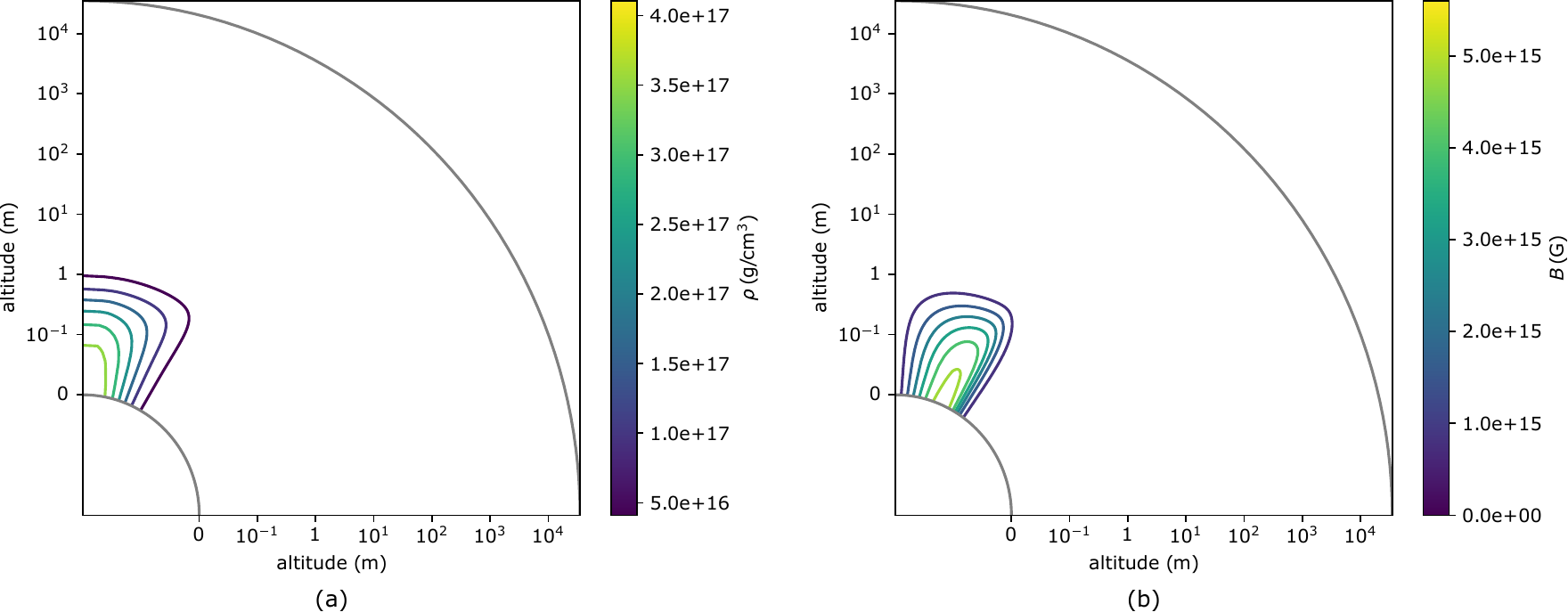}
    \caption{Hydromagnetic structure of the general relativistic equilibrium obtained for $M_a=10^{-5}M_\odot$ and $b=\psi_*/\psi_a=10$. (a) Contours of the rest-mass density $\rho$. (b) Contours of the magnitude of the magnetic field $B=\sqrt{B^2}$.}
    \label{fig:structure}
\end{figure}


\subsection{Total ellipticity}

Figure \ref{fig:structure} shows the equilibrium density and magnitude of the magnetic field of the system, plotted in meridional cross-section. We see that the mass distribution is centered at the pole of the star, as expected. In contrast, the magnitude of the magnetic field is highest around the colatitude $25^\circ$. This happens because of the flux freezing condition. The frozen-in magnetic field is dragged equatorward with the fluid, as the accreted material spreads under its own weight \citep{pm04,rossetto2023Magneticallya}. Both these configurations indicate a non-zero ellipticity of the star, as per equations \eqref{T00}--\eqref{B_mag}.

\begin{figure}[h]
    \centering
    \includegraphics[width=0.9\textwidth]{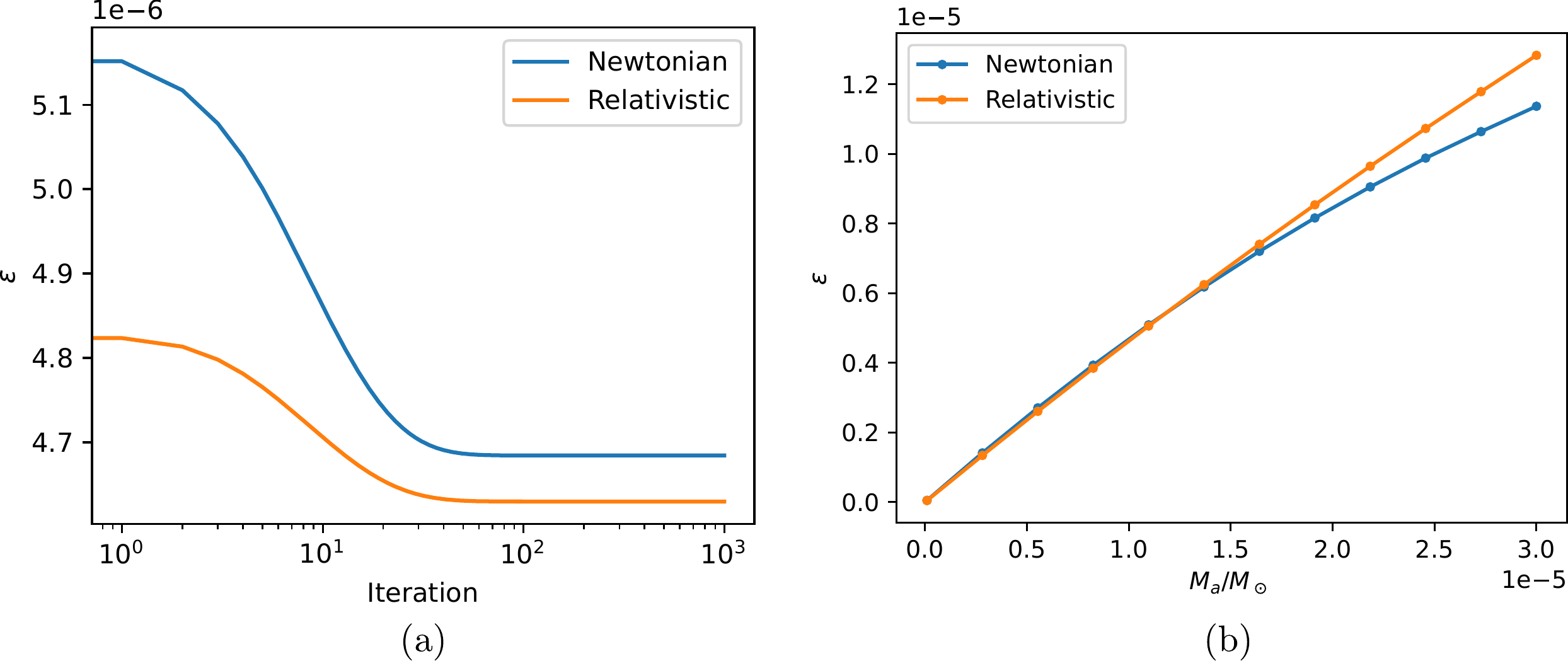}
    \caption{(a) Numerical convergence for $M_{\rm a}=10^{-5}M_\odot$ and $b=\psi_*/\psi_a=3$: total ellipticity versus iteration number of the numerical scheme. (b) Total ellipticity versus accreted mass in the range $10^{-7}\leq M_{\rm a}/M_\odot \leq 3\times10^{-5}$, for $b=\psi_*/\psi_a=3$, in the Newtonian and relativistic scenarios. In both panels, the blue and orange curves correspond to the Newtonian and relativistic calculations respectively.}
    \label{fig:total_ellip}
\end{figure}

Figure \ref{fig:total_ellip} displays calculations of the total ellipticity $\epsilon$ of the star. Panel \ref{fig:total_ellip}(a) shows the convergence achieved by the numerical solver in both the relativistic and Newtonian scenarios. The ellipticity is plotted as a function of iteration number, not as a function of time; the reader is reminded that the Grad-Shafranov equation applies in the steady state. We observe that $\epsilon$ relaxes over $\sim 50$ iterations for both scenarios, decreasing by $\approx 9\%$ and $\approx 5\%$ during the iterative process in the Newtonian and relativistic cases respectively. 

Figure \ref{fig:total_ellip}(b) shows the converged ellipticity for different values of accreted mass in the range $1\times10^{-7}\leq M_{\rm a}/M_\odot \leq 3\times10^{-5}$. The upper limit of this mass range is set by the convergence of the numerical solver. For $M_{\rm a} \gtrsim 10^{-5} M_\odot$, closed magnetic bubbles form, whose field lines do not connect to the stellar surface, and this phenomenon hinders the convergence of the solution. In the Newtonian regime, this happens for $M_{\rm a}/M_\odot\gtrsim10^{-4}b$ \citep{pm04}. In the relativistic regime, it happens for $M_{\rm a}/M_\odot\gtrsim3\times10^{-4}b$. This difference is consistent with smaller magnetic deformation in the relativistic case, as is apparent visually in Figure \ref{fig:psi}(a), where the solid curves are less deformed.  For $1\times 10^{-7} \leq M_{\rm a} / M_\odot \leq 3 \times 10^{-5}$, a simple linear regression on the curves shown in Figure \ref{fig:total_ellip}(b) yields
\begin{align}
    \epsilon_{\rm N} = 0.38M_{\rm a}/M_\odot+6\times10^{-7},
    \label{ellip_N}
    \\
    \epsilon_{\rm GR} = 0.43M_{\rm a}/M_\odot+2\times10^{-7},
    \label{ellip_GR}
\end{align}
where $\epsilon_{\rm N}$ is the Newtonian ellipticity\footnote{This agrees with Newtonian values in the literature, cf. Figure 2(a) in \cite{melatos2005Gravitational}.}, displayed in blue in Figure \ref{fig:total_ellip}, and $\epsilon_{\rm GR}$ is the relativistic ellipticitiy, displayed in orange. Equations \eqref{ellip_N} and \eqref{ellip_GR} reflect the higher ellipticity in the general relativistic case. The simple regressions \eqref{ellip_N} and \eqref{ellip_GR} must be modified for $M_{\rm a} \geq 3 \times 10^{-5} M_\odot$, when the magnetic burial process saturates; cf. Equation (8) of \cite{melatos2005Gravitational}. For $10^{-7}\leq M_{\rm a}/M_\odot \lesssim 1.3\times10^{-5}$, the relativistic and Newtonian ellipticities are similar; one finds $|\epsilon_{\rm N} - \epsilon_{\rm GR} | \leq 9.94\times10^{-8}$ throughout this range. For $M_{\rm a}\gtrsim1.3\times10^{-5} M_\odot$, the results start to diverge with the relativistic ellipticity being noticeably greater than the Newtonian one. In fractional terms, the relativistic result is $|\epsilon_{\rm GR} - \epsilon_{\rm N}| / \epsilon_{\rm N} \lesssim 12\%$ greater than its Newtonian counterpart.

\subsection{Physical interpretation}
\label{sec:phys_interp}

As discussed in Section \ref{sec:model}, general relativity reduces the deformation of the magnetic field lines, when compared to Newtonian theory (cf. Figure \ref{fig:psi}(a)). Together with flux-freezing, this implies that the density profile is narrower in the $\theta$ direction in the relativistic case. Near the polar surface, the mountain is also more compressed in the radial direction in the relativistic case. Therefore, one can expect the relativistic ellipticity to be greater than the Newtonian one, in accord with Figure \ref{fig:total_ellip}(b). 

\begin{figure}
    \centering
    \includegraphics[width=0.9\textwidth]{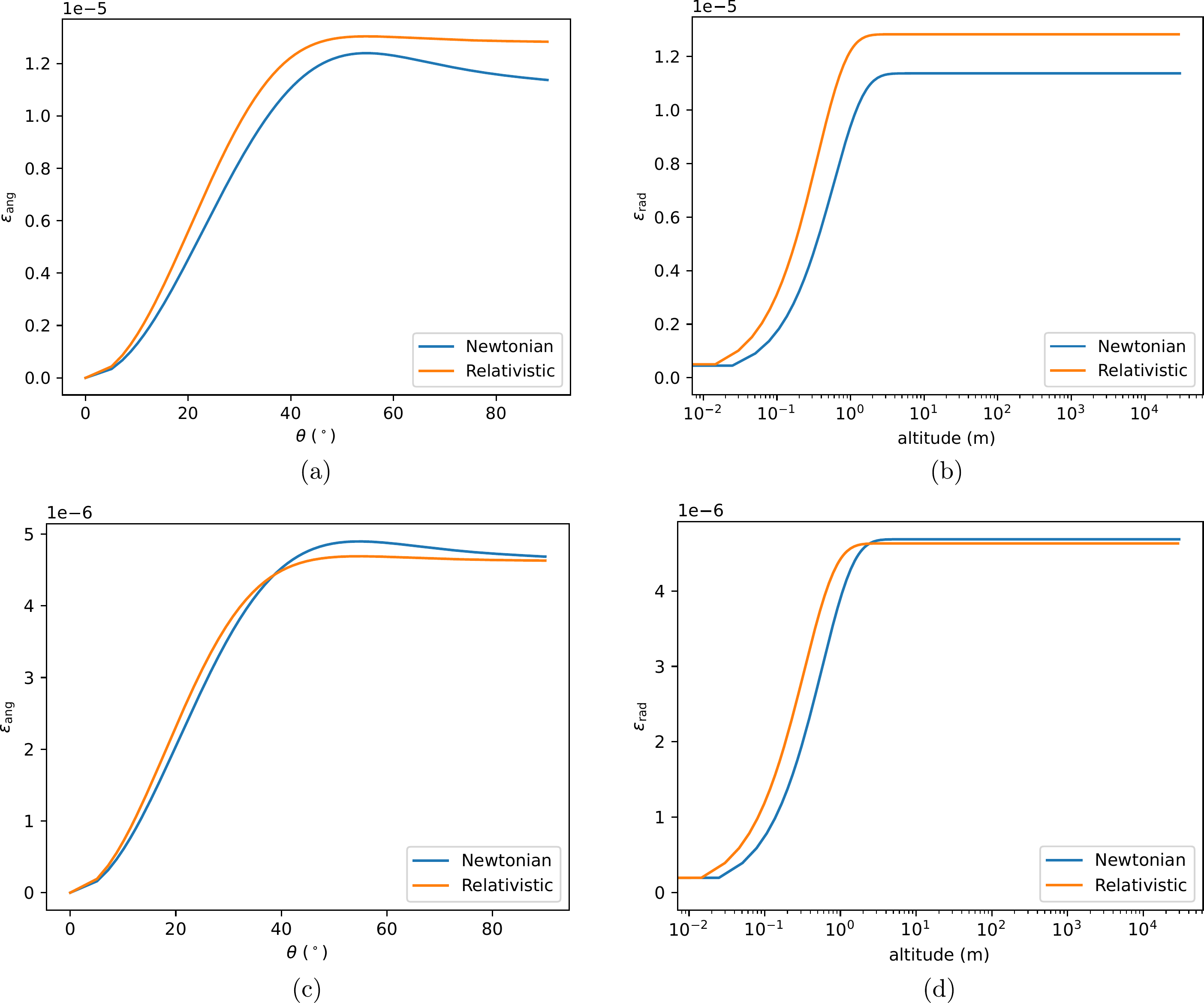}
    \caption{(a) Cumulative ellipticity $\epsilon_{\rm ang}(\theta)$ for $M_{\rm a} = 3\times10^{-5}M_\odot$. (b) Cumulative ellipticity $\epsilon_{\rm rad}(r)$ for $M_{\rm a} = 3\times10^{-5}M_\odot$. (c) Cumulative ellipticity $\epsilon_{\rm ang}(\theta)$ for $M_{\rm a} = 10^{-5}M_\odot$. (d) Cumulative ellipticity $\epsilon_{\rm rad}(r)$ for $M_{\rm a} = 10^{-5}M_\odot$. In all panels, we have $b=3$, and the blue and orange curves correspond to the Newtonian and relativistic calculations respectively.}
    \label{fig:cumulative_ellip}
\end{figure}

To analyse the angular and radial behaviours of the ellipticity, we plot in Figure \ref{fig:cumulative_ellip} the Newtonian and relativistic cumulative ellipticities $\epsilon_{\rm ang}(\theta)$ and $\epsilon_{\rm rad}(r)$, defined by
\begin{align}
    \label{cumulative_ellips}
    \epsilon_{\rm ang}(\theta) &= \frac{\pi}{I_0}\int_{0}^{\theta}\int_{R_*}^{R_{\rm m}}T^{00} r^4\sin\theta(3\cos^2\theta-1)\dd{r}\dd{\theta}\text{, }
    \\
    \epsilon_{\rm rad}(r) &= \frac{\pi}{I_0}\int_{R_*}^{r}\int_{0}^{\pi/2}T^{00} r^4\sin\theta(3\cos^2\theta-1)\dd{\theta}\dd{r},
\end{align}
where $R_{\rm m}$ is the maximum radius of the simulation box. We choose two values of accreted mass, $M_{\rm a}$, specifically $1\times10^{-5}M_\odot$ and $3\times10^{-5}M_\odot$, both with $b=3$. In Figures \ref{fig:cumulative_ellip}(a) and \ref{fig:cumulative_ellip}(c), the Newtonian ellipticity reaches a maximum at $\theta \approx 54^\circ$. This happens because, for $\theta>\arccos(3^{-1/2})$, the density contributes negatively to the ellipticity, according to equation \eqref{cumulative_ellips}. This feature is not as noticeable in the relativistic case because the mass(-energy) density for $\theta>\arccos(3^{-1/2})$ is negligible. Figures \ref{fig:cumulative_ellip}(b) and \ref{fig:cumulative_ellip}(d) show that the ellipticity, both in the relativistic and Newtonian calculations reaches its asymptotic value for $r-R_*\gtrsim1\,\rm{m}$, as the screening currents are negligible for $r-R_* \gtrsim 1 \, {\rm m}$, consistent with the calculations in Section 4 in \cite{pm04}.

The four panels in Figure \ref{fig:cumulative_ellip} show that the cumulative relativistic ellipticities have a ``head start'' compared to the cumulative Newtonian ellipticities. This is consistent with the fact that the mountain is more tightly confined in the relativistic case. However, the asymptotic cumulative relativistic ellipticity can be either greater or smaller than the Newtonian ellipticity. For the case presented in panels \ref{fig:cumulative_ellip}(a) and \ref{fig:cumulative_ellip}(b), i.e. for $M_{\rm a}=3\times10^{-5}M_\odot$, the relativistic ellipticity is greater than the Newtonian one for all radii and all angles. On the other hand, in panels \ref{fig:cumulative_ellip}(c) and \ref{fig:cumulative_ellip}(d) ($M_{\rm a}=10^{-5}M_\odot$), the Newtonian ellipticity becomes slightly greater than the relativistic for $\theta \gtrsim 38^\circ$ and $r \gtrsim 1{\rm m}$. Overall, it is hard to give an elementary physical interpretation of the change in ellipticity because of the multifactorial dependence of equation \eqref{epsilon} on the problem's variables. On the one hand, compressing the mountain in the angular direction increases the ellipticity because it minimizes negative contributions from $\sin\theta(3\cos^2\theta-1)$. On the other hand, over-compression reduces the ellipticity. Increasing the density near the pole helps to increase the ellipticity, dominated by the factor $r^4$ in equation \eqref{epsilon}.

\pagebreak

\subsection{Purely magnetic ellipticity}
\label{sec:mag_ellip}

\begin{figure}[h]
    \centering
    \includegraphics[width=0.45\textwidth]{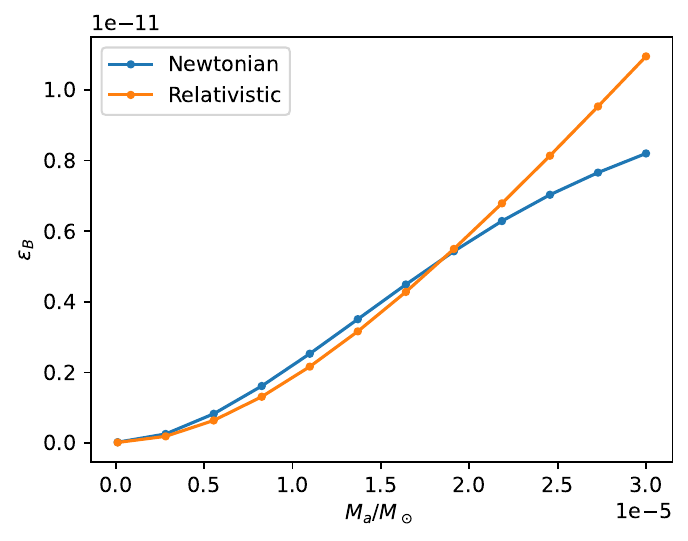}
    \caption{Direct contribution of the magnetic field to the ellipticity of the star. Magnetic ellipticity $\epsilon_B$ versus accreted mass in the range $10^{-7}\leq M_{\rm a}/M_\odot \leq 3\times10^{-5}$, for $b=\psi_*/\psi_a=3$, in the Newtonian (blue curve) and relativistic (orange curve) scenarios.}
    \label{fig:mag_ellip}
\end{figure}

The magnetic field contributes directly to the quadrupole moment \eqref{quad_mom} and the ellipticity \eqref{epsilon} of the star through the $B^2$ term in equation \eqref{T00}. This contribution is distinct from the magnetically induced nonaxisymmetry in the accreted matter, which enters through the term $\rho$ in \eqref{T00}. In the case of magnetically confined mountains, the accreted matter deforms the magnetic field and pushes it equatorwards \citep{pm04,rossetto2023Magneticallya}, increasing the magnetic pressure $B^2/2$ locally. Figure \ref{fig:mag_ellip} displays how the magnetic ellipticity increases with $M_{\rm a}$. We find that the ellipticity generated directly by the magnetic field is of the order $10^{-11}$, which is $\sim10^{6}$ times smaller than the ellipticity generated by the accreted mass (cf. Figure \ref{fig:total_ellip}). 

Recent studies \citep{hacyan2017Gravitational, nazari2020Gravitational, contopoulos2023Gravitational} investigate the gravitational radiation emitted by a purely dipolar bar magnet rotating in vacuo about an axis inclined to its magnetic axis. They find that the $B^2$ contribution is generally low but may be detectable by future-generation gravitational-wave detectors in the case of highly-magnetised fast-spinning neutron stars. In our simulations, we can recover the magnetic dipole solution in vacuo by taking the accreted mass to be low ($M_{\rm a}\sim10^{-9}M_\odot$). In the latter regime, we obtain $\epsilon_B\sim10^{-15}$. This value is consistent with the literature,  cf. equation (13) of \cite{hacyan2017Gravitational}, equation (46) of \cite{nazari2020Gravitational} for $B\sim10^{12}\, \rm{G}$, and equation (57) of \cite{contopoulos2023Gravitational}.

\pagebreak

\subsection{Compactness}

\begin{figure}[h]
    \centering
    \includegraphics[width=0.95\textwidth]{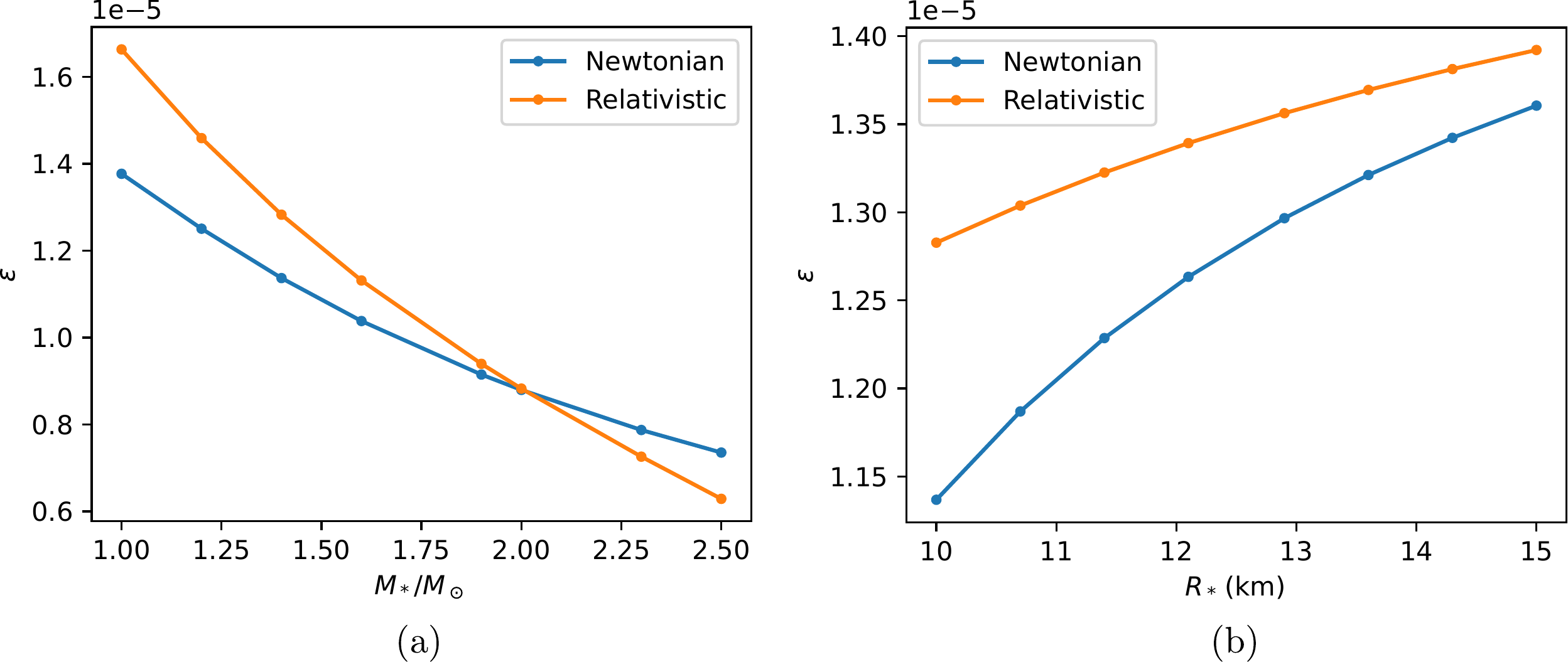}
    \caption{Effect of the compactness on the star's ellipticity for $M_{\rm a}=3\times10^{-5}M_\odot$ and $b=3$. (a) Ellipticity versus neutron star mass in the range $1\leq M_*/M_\odot\leq2.5$ for $R_*=10\,\mathrm{km}$ fixed. (b) Ellipticity versus neutron star radius in the range $10\,\mathrm{km}\leq R_*\leq15\,\mathrm{km}$ for $M_*=1.4M\odot$ fixed. In both panels, the blue and orange curves correspond to the Newtonian and relativistic calculations respectively.}
    \label{fig:compact}
\end{figure}

The relativistic corrections to the mountain structure depend on the compactness $M_*/R_*$ of the star. These corrections are greater for more compact objects. Figure \ref{fig:compact} shows the effect of the compactness of the star for $M_{\rm a} = 3\times10^{-5}M_\odot$ and $b=3$. Figure \ref{fig:compact}(a) shows the effects of varying $M_*$ in the range $1\leq M_*/M_\odot\leq2.5$ with $R_*=10\,\mathrm{km}$ fixed, i.e.  compactness in the range $0.15\leq M_*/R_* \leq0.37$. The percentage difference varies from $\approx -20\%$ to $\approx 14\%$ for the range of $M_*$ presented.

Figure \ref{fig:compact}(b) shows the effects of varying $R_*$ in the range $10\,\mathrm{km}\leq R_*\leq15\,\mathrm{km}$ with $M_*=1.4M_\odot$ fixed, i.e. compactness in the range $0.20\leq M_*/R_* \leq0.31$. Figure \ref{fig:compact}(b) shows the difference between the Newtonian and relativistic results decreases as the radius increases (decrease in compactness). The results get closer, going from a percentage difference of $\approx 12\%$ for $R_*=10\,\mathrm{km}$ to $\approx 2\%$ for $R_*=15\,\mathrm{km}$.

%% file: Sections/4-discussion.tex
The present work focuses on the modifications arising from including general relativity in models of the formation of magnetically confined mountains and hence the emission of continuous gravitational waves by accreting neutron stars. The baseline for comparison is the Newtonian model calculated by \cite{melatos2005Gravitational} for an isothermal equation of state. The modifications brought by general relativity to the hydromagnetic structure of the mountain are discussed in detail by \cite{rossetto2023Magneticallya}. The main effect is to increase the ellipticity in the relativistic case, as can be seen in equations \eqref{ellip_N} and \eqref{ellip_GR}. 

The characteristic gravitational wave strain $h_0$ from a star with a static mountain with ellipticity $\epsilon$ can be written as
\begin{equation}
    \label{strain_estimate}
    h_0 = 4.7\times10^{-26}
    \paren{\frac{\epsilon}{10^{-6}}}
    \paren{\frac{I_0}{1.1\times10^{38}\mathrm{kg\,m^2}}}
    \paren{\frac{f}{100\,\rm{Hz}}}^2
    \paren{\frac{1\,\rm{kpc}}{r}}.
\end{equation}
From the ellipticity results shown in Figure \ref{fig:total_ellip} we obtain strains between $\sim 10^{-26}$ and $\sim 10^{-25}$. The relativistic results are up to $12\%$ greater than the Newtonian ones, depending on $M_{\rm a}$. The purely magnetic contribution from $\epsilon_B$ is negligible, generating a wave strain of the order $10^{-32}$.


Several important effects already modeled in the Newtonian literature must also be incorporated into the relativistic model. A summary of these effects can be found in Section \ref{sec:intro}. The unmodeled effect that probably affects the system the most is the choice of the equation of state. In the Newtonian literature \citep{priymak2011Quadrupole}, a polytropic or piecewise-polytropic equation of state reduces the quadrupole between three and four orders of magnitude compared to an isothermal equation of state. Furthermore, dynamical effects need to be included. The Grad-Shafranov equation cannot model these effects, as it is an equilibrium equation. Dynamical effects control the stability of the accreted mountain \citep{payne2007Burial, vigelius2008Threedimensional,mukherjee2013MHD,mukherjee2017Revisiting}.

%% file: Sections/acknowledgements.tex
This work received financial support from the Australian Research Council Centre of Excellence for Gravitational Wave Discovery (OzGrav), through project number CE170100004. This work was partially supported financially by the Catalyst Fund provided by the New Zealand Ministry of Business, Innovation and Employment and administered by the Royal Society Te Apārangi; and the Division of Science of the University of Otago, New Zealand.

%% file: main.bbl
\begin{thebibliography}{}
\expandafter\ifx\csname natexlab\endcsname\relax\def\natexlab#1{#1}\fi
\providecommand{\url}[1]{\href{#1}{#1}}
\providecommand{\dodoi}[1]{doi:~\href{http://doi.org/#1}{\nolinkurl{#1}}}
\providecommand{\doeprint}[1]{\href{http://ascl.net/#1}{\nolinkurl{http://ascl.net/#1}}}
\providecommand{\doarXiv}[1]{\href{https://arxiv.org/abs/#1}{\nolinkurl{https://arxiv.org/abs/#1}}}

\bibitem[{Abbott(2022)}]{abbott2022Modelbased}
Abbott, R. 2022, The Astrophysical Journal Letters

\bibitem[{Abbott {et~al.}(2021)Abbott, Abbott, Abraham, Acernese, Ackley, Adams, Adams, Adhikari, Adya, Affeldt, Agarwal, Agathos, Agatsuma, Aggarwal, Aguiar, Aiello, Ain, Ajith, Akutsu, Aleman, Allen, Allocca, Altin, Amato, Anand, Ananyeva, Anderson, Anderson, Ando, Angelova, Ansoldi, Antelis, Antier, Appert, Arai, Arai, Arai, Araki, Araya, Araya, Areeda, Ar{\`e}ne, Aritomi, Arnaud, Aronson, Arun, Asada, Asali, Ashton, Aso, Aston, Astone, Aubin, Aufmuth, AultONeal, Austin, Babak, Badaracco, Bader, Bae, Bae, Baer, Bagnasco, Bai, Baiotti, Baird, Bajpai, Ball, Ballardin, Ballmer, Bals, Balsamo, Baltus, Banagiri, Bankar, Bankar, Barayoga, Barbieri, Barish, Barker, Barneo, Barone, Barr, Barsotti, Barsuglia, Barta, Bartlett, Barton, Bartos, Bassiri, Basti, Bawaj, Bayley, Baylor, Bazzan, B{\'e}csy, Bedakihale, Bejger, Belahcene, Benedetto, Beniwal, Benjamin, Bennett, Bentley, BenYaala, Bergamin, Berger, Bernuzzi, Bersanetti, Bertolini, Betzwieser, Bhandare, Bhandari, Bhattacharjee, Bhaumik, Bidler, Bilenko,
  Billingsley, Birney, Birnholtz, Biscans, Bischi, Biscoveanu, Bisht, Biswas, Bitossi, Bizouard, Blackburn, Blackman, Blair, Blair, Blair, Bobba, Bode, Boer, Bogaert, Boldrini, Bondu, Bonilla, Bonnand, Booker, Boom, Bork, Boschi, Bose, Bose, Bossilkov, Boudart, Bouffanais, Bozzi, Bradaschia, Brady, Bramley, Branch, Branchesi, Brau, Breschi, Briant, Briggs, Brillet, Brinkmann, Brockill, Brooks, Brooks, Brown, Brunett, Bruno, Bruntz, Bryant, Buikema, Bulik, Bulten, Buonanno, Buscicchio, Buskulic, Byer, Cadonati, Caesar, Cagnoli, Cahillane, Cain~Iii, Calder{\'o}n~Bustillo, Callaghan, Callister, Calloni, Camp, Canepa, Cannavacciuolo, Cannon, Cao, Cao, Cao, Capocasa, Capote, Carapella, Carbognani, Carlin, Carney, Carpinelli, Carullo, Carver, Casanueva~Diaz, Casentini, Castaldi, Caudill, Cavagli{\`a}, Cavalier, Cavalieri, Cella, {Cerd{\'a}-Dur{\'a}n}, Cesarini, Chaibi, Chakravarti, Champion, Chan, Chan, Chan, Chan, Chandra, Chanial, Chao, Charlton, Chase, {Chassande-Mottin}, Chatterjee, Chaturvedi, Chen, Chen,
  Chen, Chen, Chen, Chen, Chen, Chen, Chen, Cheng, Cheong, Cheung, Chia, Chiadini, Chiang, Chierici, Chincarini, Chiofalo, Chiummo, Cho, Cho, Choate, Choudhary, Choudhary, Christensen, Chu, Chu, Chu, Chua, Chung, Ciani, Ciecielag, Cie{\'s}lar, Cifaldi, Ciobanu, Ciolfi, Cipriano, Cirone, Clara, Clark, Clark, Clarke, Clearwater, Clesse, Cleva, Coccia, Cohadon, Cohen, Cohen, Colleoni, Collette, Colpi, Compton, Constancio~Jr., Conti, Cooper, Corban, Corbitt, {Cordero-Carri{\'o}n}, Corezzi, Corley, Cornish, Corre, Corsi, Cortese, Costa, Cotesta, Coughlin, Coughlin, Coulon, Countryman, Cousins, Couvares, Covas, Coward, Cowart, Coyne, Coyne, Creighton, Creighton, Criswell, Croquette, Crowder, Cudell, Cullen, Cumming, Cummings, Cuoco, Cury{\l}o, Dal~Canton, D{\'a}lya, Dana, DaneshgaranBajastani, D'Angelo, Danilishin, D'Antonio, Danzmann, {Darsow-Fromm}, Dasgupta, Datrier, Dattilo, Dave, Davier, Davies, Davis, Daw, Dean, DeBra, Deenadayalan, Degallaix, De~Laurentis, Del{\'e}glise, Del~Favero, De~Lillo, De~Lillo,
  Del~Pozzo, DeMarchi, De~Matteis, D'Emilio, Demos, Dent, Depasse, De~Pietri, De~Rosa, De~Rossi, DeSalvo, De~Simone, Dhurandhar, D{\'i}az, {Diaz-Ortiz Jr.}, Didio, Dietrich, Di~Fiore, Di~Fronzo, Di~Giorgio, Di~Giovanni, Di~Girolamo, Di~Lieto, Ding, Di~Pace, Di~Palma, Di~Renzo, Divakarla, Dmitriev, Doctor, D'Onofrio, Donovan, Dooley, Doravari, Dorrington, Drago, Driggers, Drori, Du, Ducoin, Dupej, Durante, D'Urso, Duverne, Dwyer, Easter, Ebersold, Eddolls, Edelman, Edo, Edy, Effler, Eguchi, Eichholz, Eikenberry, Eisenmann, Eisenstein, Ejlli, Enomoto, Errico, Essick, Estell{\'e}s, Estevez, Etienne, Etzel, Evans, Evans, Ewing, Fafone, Fair, Fairhurst, Fan, Farah, Farinon, Farr, Farr, Farrow, {Fauchon-Jones}, Favata, Fays, Fazio, Feicht, Fejer, Feng, Fenyvesi, Ferguson, {Fernandez-Galiana}, Ferrante, Ferreira, Fidecaro, Figura, Fiori, Fishbach, Fisher, Fittipaldi, Fiumara, Flaminio, Floden, Flynn, Fong, Font, Fornal, Forsyth, Franke, Frasca, Frasconi, Frederick, Frei, Freise, Frey, Fritschel, Frolov, Fronz{\'e},
  Fujii, Fujikawa, Fukunaga, Fukushima, Fulda, Fyffe, Gabbard, Gadre, Gaebel, Gair, Gais, Galaudage, Gamba, Ganapathy, Ganguly, Gao, Gaonkar, Garaventa, {Garc{\'i}a-N{\'u}{\~n}ez}, {Garc{\'i}a-Quir{\'o}s}, Garufi, Gateley, Gaudio, Gayathri, Ge, Gemme, Gennai, George, Gergely, Gewecke, Ghonge, Ghosh, Ghosh, Ghosh, Ghosh, Ghosh, Giacomazzo, Giacoppo, Giaime, Giardina, Gibson, Gier, Giesler, Giri, Gissi, Glanzer, Gleckl, Godwin, Goetz, Goetz, Gohlke, Goncharov, Gonz{\'a}lez, Gopakumar, Gosselin, Gouaty, Grace, Grado, Granata, Granata, Grant, Gras, Grassia, Gray, Gray, Greco, Green, Green, Gretarsson, Gretarsson, Griffith, Griffiths, Griggs, Grignani, Grimaldi, Grimes, Grimm, Grote, Grunewald, Gruning, Guerrero, Guidi, Guimaraes, Guix{\'e}, Gulati, Guo, Guo, Gupta, Gupta, Gupta, Gustafson, Gustafson, Guzman, Ha, Haegel, Hagiwara, Haino, Halim, Hall, Hamilton, Hammond, Han, Haney, Hanks, Hanna, Hannam, Hannuksela, Hansen, Hansen, Hanson, Harder, Hardwick, Haris, Harms, Harry, Harry, Hartwig, Hasegawa, Haskell,
  Hasskew, Haster, Hattori, Haughian, Hayakawa, Hayama, Hayes, Healy, Heidmann, Heintze, Heinze, Heinzel, Heitmann, Hellman, Hello, {Helmling-Cornell}, Hemming, Hendry, Heng, Hennes, Hennig, Hennig, Hernandez~Vivanco, Heurs, Hild, Hill, Himemoto, Hines, Hiranuma, Hirata, Hirose, Hochheim, Hofman, Hohmann, Holgado, Holland, Hollows, Holmes, Holt, Holz, Hong, Hopkins, Hough, Howell, Hoy, Hoyland, Hreibi, Hsieh, Hsu, Huang, Huang, Huang, Huang, Huang, Huang, H{\"u}bner, Huddart, Huerta, Hughey, Hui, Hui, Husa, Huttner, Huxford, {Huynh-Dinh}, Ide, Idzkowski, Iess, Ikenoue, Imam, Inayoshi, Inchauspe, Ingram, Inoue, Intini, Ioka, Isi, Isleif, Ito, Itoh, Iyer, Izumi, JaberianHamedan, Jacqmin, Jadhav, Jadhav, James, Jan, Jani, Janssens, Janthalur, Jaranowski, Jariwala, Jaume, Jenkins, Jeon, Jeunon, Jia, Jiang, Jin, Johns, Jones, Jones, Jones, Jones, Jones, Jonker, Ju, Jung, Jung, Junker, Kaihotsu, Kajita, Kakizaki, Kalaghatgi, Kalogera, Kamai, Kamiizumi, Kanda, Kandhasamy, Kang, Kanner, Kao, Kapadia, Kapasi, Karat,
  Karathanasis, Karki, Kashyap, Kasprzack, Kastaun, Katsanevas, Katsavounidis, Katzman, Kaur, Kawabe, Kawaguchi, Kawai, Kawasaki, K{\'e}f{\'e}lian, Keitel, Key, Khadka, Khalili, Khan, Khan, Khazanov, Khetan, Khursheed, Kijbunchoo, Kim, Kim, Kim, Kim, Kim, Kim, Kimball, Kimura, King, {Kinley-Hanlon}, Kirchhoff, Kissel, Kita, Kitazawa, Kleybolte, Klimenko, Knee, Knowles, Knyazev, Koch, Koekoek, Kojima, Kokeyama, Koley, Kolitsidou, Kolstein, Komori, Kondrashov, Kong, Kontos, Koper, Korobko, Kotake, Kovalam, Kozak, Kozakai, Kozu, Kringel, Krishnendu, Kr{\'o}lak, Kuehn, Kuei, Kumar, Kumar, Kumar, Kumar, Kume, Kuns, Kuo, Kuo, Kuromiya, Kuroyanagi, Kusayanagi, Kwak, Kwang, Laghi, Lalande, Lam, Lamberts, Landry, Lane, Lang, Lange, Lantz, La~Rosa, {Lartaux-Vollard}, Lasky, Laxen, Lazzarini, Lazzaro, Leaci, Leavey, Lecoeuche, Lee, Lee, Lee, Lee, Lee, Lee, Lehmann, Lema{\^i}tre, Leon, Leonardi, Leroy, Letendre, Levin, Leviton, Li, Li, Li, Li, Li, Li, Lin, Lin, Lin, Lin, Lin, Linde, Linker, Linley, Littenberg, Liu, Liu,
  Liu, Liu, {Llorens-Monteagudo}, Lo, Lockwood, Lollie, London, Longo, Lopez, Lorenzini, Loriette, Lormand, Losurdo, Lough, Lousto, Lovelace, L{\"u}ck, Lumaca, Lundgren, Luo, Macas, MacInnis, Macleod, MacMillan, Macquet, Maga{\~n}a~Hernandez, {Maga{\~n}a-Sandoval}, Magazz{\`u}, Magee, Maggiore, Majorana, Makarem, Maksimovic, Maliakal, Malik, Man, Mandic, Mangano, Mango, Mansell, Manske, Mantovani, Mapelli, Marchesoni, Marchio, Marion, Mark, M{\'a}rka, M{\'a}rka, Markakis, Markosyan, Markowitz, Maros, Marquina, Marsat, Martelli, Martin, Martin, Martinez, Martinez, Martinovic, Martynov, Marx, Masalehdan, Mason, Massera, Masserot, Massinger, {Masso-Reid}, Mastrogiovanni, Matas, {Mateu-Lucena}, Matichard, Matiushechkina, Mavalvala, McCann, McCarthy, McClelland, McClincy, McCormick, McCuller, McGhee, McGuire, McIsaac, McIver, McManus, McRae, McWilliams, Meacher, Mehmet, Mehta, Melatos, Melchor, Mendell, {Menendez-Vazquez}, Menoni, Mercer, Mereni, Merfeld, Merilh, Merritt, Merzougui, Meshkov, Messenger, Messick,
  Meyers, Meylahn, Mhaske, Miani, Miao, Michaloliakos, Michel, Michimura, Middleton, Milano, Miller, Millhouse, Mills, Milotti, {Milovich-Goff}, Minazzoli, Minenkov, Mio, Mir, Mishkin, Mishra, Mishra, Mistry, Mitra, Mitrofanov, Mitselmakher, Mittleman, Miyakawa, Miyamoto, Miyazaki, Miyo, Miyoki, Mo, Mogushi, Mohapatra, Mohite, Molina, {Molina-Ruiz}, Mondin, Montani, Moore, Moraru, Morawski, More, Moreno, Moreno, Mori, Morisaki, Moriwaki, Mours, {Mow-Lowry}, Mozzon, Muciaccia, Mukherjee, Mukherjee, Mukherjee, Mukherjee, Mukund, Mullavey, Munch, Mu{\~n}iz, Murray, Musenich, Nadji, Nagano, Nagano, Nagar, Nakamura, Nakano, Nakano, Nakashima, Nakayama, Nardecchia, Narikawa, Naticchioni, Nayak, Nayak, Negishi, Neil, Neilson, Nelemans, Nelson, Nery, Neunzert, Ng, Ng, Nguyen, Nguyen, Nguyen, Nguyen~Quynh, Ni, Nichols, Nishizawa, Nissanke, Nocera, Noh, Norman, North, Nozaki, Nuttall, Oberling, O'Brien, Obuchi, O'Dell, Ogaki, Oganesyan, Oh, Oh, Oh, Ohashi, Ohishi, Ohkawa, Ohme, Ohta, Okada, Okutani, Okutomi, Olivetto,
  Oohara, Ooi, Oram, O'Reilly, Ormiston, Ormsby, Ortega, O'Shaughnessy, O'Shea, Oshino, Ossokine, Osthelder, Otabe, Ottaway, Overmier, Pace, Pagano, Page, Pagliaroli, Pai, Pai, Palamos, Palashov, Palomba, Pan, Panda, Pang, Pang, Pankow, Pannarale, Pant, Paoletti, Paoli, Paolone, Parisi, Park, Parker, Pascucci, Pasqualetti, Passaquieti, Passuello, Patel, Patricelli, Payne, Pechsiri, Pedraza, Pegoraro, Pele, Pe{\~n}a~Arellano, Penn, Perego, Pereira, Pereira, Perez, P{\'e}rigois, Perreca, Perri{\`e}s, Petermann, Petterson, Pfeiffer, Pham, Phukon, Piccinni, Pichot, Piendibene, Piergiovanni, Pierini, Pierro, Pillant, Pilo, Pinard, Pinto, Piotrzkowski, Piotrzkowski, Pirello, Pitkin, Placidi, Plastino, Pluchar, Poggiani, Polini, Pong, Ponrathnam, Popolizio, Porter, Powell, Pracchia, Pradier, Prajapati, Prasai, Prasanna, Pratten, Prestegard, Principe, Prodi, Prokhorov, Prosposito, Prudenzi, Puecher, Punturo, Puosi, Puppo, P{\"u}rrer, Qi, Quetschke, Quinonez, {Quitzow-James}, Raab, Raaijmakers, Radkins, Radulesco,
  Raffai, Rail, Raja, Rajan, Ramirez, Ramirez, {Ramos-Buades}, Rana, Rapagnani, Rapol, Ratto, Raymond, Raza, Razzano, Read, Rees, Regimbau, Rei, Reid, Reitze, Relton, Rettegno, Ricci, Richardson, Richardson, Richardson, Ricker, Riemenschneider, Riles, Rizzo, Robertson, Robie, Robinet, Rocchi, Rocha, Rodriguez, {Rodriguez-Soto}, Rolland, Rollins, Roma, Romanelli, Romano, Romel, Romero, {Romero-Shaw}, Romie, Rose, Rosi{\'n}ska, Rosofsky, Ross, Rowan, Rowlinson, Roy, Roy, Rozza, Ruggi, Ryan, Sachdev, Sadecki, Sadiq, Sago, Saito, Saito, Sakai, Sakai, Sakellariadou, Sakuno, Salafia, Salconi, Saleem, Salemi, Samajdar, Sanchez, Sanchez, Sanchez, {Sanchis-Gual}, Sanders, Sanuy, Saravanan, Sarin, Sassolas, Satari, Sathyaprakash, Sato, Sato, Sauter, Savage, Savant, Sawada, Sawant, Sawant, Sayah, Schaetzl, Scheel, Scheuer, {Schindler-Tyka}, Schmidt, Schnabel, Schneewind, Schofield, Sch{\"o}nbeck, Schulte, Schutz, Schwartz, Scott, Scott, {Seglar-Arroyo}, Seidel, Sekiguchi, Sekiguchi, Sellers, Sengupta, Sennett, Sentenac,
  Seo, Sequino, Sergeev, Setyawati, Shaffer, Shahriar, Shams, Shao, Sharifi, Sharma, Sharma, Shawhan, Shcheblanov, Shen, Shibagaki, Shikauchi, Shimizu, Shimoda, Shimode, Shink, Shinkai, Shishido, Shoda, Shoemaker, Shoemaker, Shukla, ShyamSundar, Sieniawska, Sigg, Singer, Singh, Singh, Singha, Sintes, Sipala, Skliris, Slagmolen, {Slaven-Blair}, Smetana, Smith, Smith, Somala, Somiya, Son, Soni, Soni, Sorazu, Sordini, Sorrentino, Sorrentino, Sotani, Soulard, Souradeep, Sowell, Spagnuolo, Spencer, Spera, Srivastava, Srivastava, Staats, Stachie, Steer, Steinlechner, Steinlechner, Stops, Stover, Strain, Strang, Stratta, Strunk, Sturani, Stuver, S{\"u}dbeck, Sudhagar, Sudhir, Sugimoto, Suh, Summerscales, Sun, Sun, Sunil, Sur, Suresh, Sutton, Suzuki, Suzuki, Swinkels, Szczepa{\'n}czyk, Szewczyk, Tacca, Tagoshi, Tait, Takahashi, Takahashi, Takamori, Takano, Takeda, Takeda, Talbot, Tanaka, Tanaka, Tanaka, Tanaka, Tanaka, Tanasijczuk, Tanioka, Tanner, Tao, Tapia, Tapia San~Martin, Tapia San~Martin, Tasson, Telada,
  Tenorio, Terkowski, Test, Thirugnanasambandam, Thomas, Thomas, Thompson, Thondapu, Thorne, Thrane, Tiwari, Tiwari, Tiwari, Toland, Tolley, Tomaru, Tomigami, Tomura, Tonelli, {Torres-Forn{\'e}}, Torrie, Tosta E~Melo, T{\"o}yr{\"a}, Trapananti, Travasso, Traylor, Tringali, Tripathee, Troiano, Trovato, Trozzo, Trudeau, Tsai, Tsai, Tsang, Tsang, Tsao, Tse, Tso, Tsubono, Tsuchida, Tsukada, Tsuna, Tsutsui, Tsuzuki, Turconi, Tuyenbayev, Ubhi, Uchikata, Uchiyama, Udall, Ueda, Uehara, Ueno, Ueshima, Ugolini, Unnikrishnan, Uraguchi, Urban, Ushiba, Usman, Utina, Vahlbruch, Vajente, Vajpeyi, Valdes, Valentini, Valsan, Van~Bakel, Van~Beuzekom, Van Den~Brand, Van Den~Broeck, {Vander-Hyde}, Van Der~Schaaf, Van~Heijningen, Vanosky, Van~Putten, Vardaro, Vargas, Varma, Vas{\'u}th, Vecchio, Vedovato, Veitch, Veitch, Venkateswara, Venneberg, Venugopalan, Verkindt, Verma, Veske, Vetrano, Vicer{\'e}, Viets, {Villa-Ortega}, Vinet, Vitale, Vo, Vocca, Von~Reis, Von~Wrangel, Vorvick, Vyatchanin, Wade, Wade, Wagner, Walet, Walker,
  Wallace, Wallace, Walsh, Wang, Wang, Wang, Ward, Warner, Was, Washimi, Washington, Watchi, Weaver, Wei, Weinert, Weinstein, Weiss, Weller, Wellmann, Wen, We{\ss}els, Westhouse, Wette, Whelan, White, Whiting, Whittle, Wilken, Williams, Williams, Williamson, Willis, Willke, Wilson, Winkler, Wipf, Wlodarczyk, Woan, Woehler, Wofford, Wong, Wu, Wu, Wu, Wu, Wysocki, Xiao, Xu, Yamada, Yamamoto, Yamamoto, Yamamoto, Yamamoto, Yamashita, Yamazaki, Yang, Yang, Yang, Yang, Yang, Yap, Yeeles, Yelikar, Ying, Yokogawa, Yokoyama, Yokozawa, Yoon, Yoshioka, Yu, Yu, Yuzurihara, Zadro{\.z}ny, Zanolin, Zeidler, Zelenova, Zendri, Zevin, Zhan, Zhang, Zhang, Zhang, Zhang, Zhang, Zhao, Zhao, Zhao, Zhao, Zhou, Zhu, Zhu, Zucker, \& Zweizig}]{abbott2021Searches}
Abbott, R., Abbott, T.~D., Abraham, S., {et~al.} 2021, The Astrophysical Journal, 921, 80, \dodoi{10.3847/1538-4357/ac17ea}

\bibitem[{Abbott {et~al.}(2022{\natexlab{a}})Abbott, Abbott, Acernese, Ackley, Adams, Adhikari, Adhikari, Adya, Affeldt, Agarwal, Agathos, Agatsuma, Aggarwal, Aguiar, Aiello, Ain, Ajith, Akutsu, Albanesi, Allocca, Altin, Amato, Anand, Anand, Ananyeva, Anderson, Anderson, Ando, Andrade, Andres, Andri{\'c}, Angelova, Ansoldi, Antelis, Antier, Appert, Arai, Arai, Arai, Araki, Araya, Araya, Areeda, Ar{\`e}ne, Aritomi, Arnaud, Aronson, Arun, Asada, Asali, Ashton, Aso, Assiduo, Aston, Astone, Aubin, Austin, Babak, Badaracco, Bader, Badger, Bae, Bae, Baer, Bagnasco, Bai, Bailes, Baiotti, Baird, Bajpai, Ball, Ballardin, Ballmer, Balsamo, Baltus, Banagiri, Bankar, Barayoga, Barbieri, Barish, Barker, Barneo, Barone, Barr, Barsotti, Barsuglia, Barta, Bartlett, Barton, Bartos, Bassiri, Basti, Bawaj, Bayley, Baylor, Bazzan, B{\'e}csy, Bedakihale, Bejger, Belahcene, Benedetto, Beniwal, Bennett, Bentley, BenYaala, Bergamin, Berger, Bernuzzi, Bersanetti, Bertolini, Betzwieser, Beveridge, Bhandare, Bhardwaj, Bhattacharjee,
  Bhaumik, Bilenko, Billingsley, Bini, Birney, Birnholtz, Biscans, Bischi, Biscoveanu, Bisht, Biswas, Bitossi, Bizouard, Blackburn, Blair, Blair, Blair, Bobba, Bode, Boer, Bogaert, Boldrini, Bonavena, Bondu, Bonilla, Bonnand, Booker, Boom, Bork, Boschi, Bose, Bose, Bossilkov, Boudart, Bouffanais, Bozzi, Bradaschia, Brady, Bramley, Branch, Branchesi, Brau, Breschi, Briant, Briggs, Brillet, Brinkmann, Brockill, Brooks, Brooks, Brown, Brunett, Bruno, Bruntz, Bryant, Bulik, Bulten, Buonanno, Buscicchio, Buskulic, Buy, Byer, Cadonati, Cagnoli, Cahillane, Bustillo, Callaghan, Callister, Calloni, Cameron, Camp, Canepa, Canevarolo, Cannavacciuolo, Cannon, Cao, Cao, Capocasa, Capote, Carapella, Carbognani, Carlin, Carney, Carpinelli, Carrillo, Carullo, Carver, Diaz, Casentini, Castaldi, Caudill, Cavagli{\`a}, Cavalier, Cavalieri, Ceasar, Cella, {Cerd{\'a}-Dur{\'a}n}, Cesarini, Chaibi, Chakravarti, Subrahmanya, Champion, Chan, Chan, Chan, Chan, Chan, Chandra, Chanial, Chao, Charlton, Chase, {Chassande-Mottin},
  Chatterjee, Chatterjee, Chatterjee, Chaturvedi, Chaty, Chen, Chen, Chen, Chen, Chen, Chen, Chen, Chen, Cheng, Cheong, Cheung, Chia, Chiadini, Chiang, Chiarini, Chierici, Chincarini, Chiofalo, Chiummo, Cho, Cho, Choudhary, Choudhary, Christensen, Chu, Chu, Chu, Chua, Chung, Ciani, Ciecielag, Cie{\'s}lar, Cifaldi, Ciobanu, Ciolfi, Cipriano, Cirone, Clara, Clark, Clark, Clarke, Clearwater, Clesse, Cleva, Coccia, Codazzo, Cohadon, Cohen, Cohen, Colleoni, Collette, Colombo, Colpi, Compton, Constancio, Conti, Cooper, Corban, Corbitt, {Cordero-Carri{\'o}n}, Corezzi, Corley, Cornish, Corre, Corsi, Cortese, Costa, Cotesta, Coughlin, Coulon, Countryman, Cousins, Couvares, Coward, Cowart, Coyne, Coyne, Creighton, Creighton, Criswell, Croquette, Crowder, Cudell, Cullen, Cumming, Cummings, Cunningham, Cuoco, Cury{\l}o, Dabadie, Canton, Dall'Osso, D{\'a}lya, Dana, DaneshgaranBajastani, D'Angelo, Danilishin, D'Antonio, Danzmann, {Darsow-Fromm}, Dasgupta, Datrier, Datta, Dattilo, Dave, Davier, Davies, Davis, Davis, Daw,
  Dean, DeBra, Deenadayalan, Degallaix, De~Laurentis, Del{\'e}glise, Del~Favero, De~Lillo, De~Lillo, Del~Pozzo, DeMarchi, De~Matteis, D'Emilio, Demos, Dent, Depasse, De~Pietri, De~Rosa, De~Rossi, DeSalvo, De~Simone, Dhurandhar, D{\'i}az, {Diaz-Ortiz}, Didio, Dietrich, Fiore, Fronzo, Giorgio, Giovanni, Giovanni, Girolamo, Lieto, Ding, Pace, Palma, Renzo, Divakarla, Dmitriev, Doctor, D'Onofrio, Donovan, Dooley, Doravari, Dorrington, Drago, Driggers, Drori, Ducoin, Dupej, Durante, D'Urso, Duverne, Dwyer, Eassa, Easter, Ebersold, Eckhardt, Eddolls, Edelman, Edo, Edy, Effler, Eguchi, Eichholz, Eikenberry, Eisenmann, Eisenstein, Ejlli, Engelby, Enomoto, Errico, Essick, Estell{\'e}s, Estevez, Etienne, Etzel, Evans, Evans, Ewing, Fafone, Fair, Fairhurst, Farah, Farinon, Farr, Farr, Farrow, {Fauchon-Jones}, Favaro, Favata, Fays, Fazio, Feicht, Fejer, Fenyvesi, Ferguson, {Fernandez-Galiana}, Ferrante, Ferreira, Fidecaro, Figura, Fiori, Fishbach, Fisher, Fittipaldi, Fiumara, Flaminio, Floden, Fong, Font, Fornal,
  Forsyth, Franke, Frasca, Frasconi, Frederick, Freed, Frei, Freise, Frey, Fritschel, Frolov, Fronz{\'e}, Fujii, Fujikawa, Fukunaga, Fukushima, Fulda, Fyffe, Gabbard, Gadre, Gair, Gais, Galaudage, Gamba, Ganapathy, Ganguly, Gao, Gaonkar, Garaventa, {Garc{\'i}a-N{\'u}{\~n}ez}, {Garc{\'i}a-Quir{\'o}s}, Garufi, Gateley, Gaudio, Gayathri, Ge, Gemme, Gennai, George, Gerberding, Gergely, Gewecke, Ghonge, Ghosh, Ghosh, Ghosh, Ghosh, Giacomazzo, Giacoppo, Giaime, Giardina, Gibson, Gier, Giesler, Giri, Gissi, Glanzer, Gleckl, Godwin, Goetz, Goetz, Gohlke, Goncharov, Gonz{\'a}lez, Gopakumar, Gosselin, Gouaty, Gould, Grace, Grado, Granata, Granata, Grant, Gras, Grassia, Gray, Gray, Greco, Green, Green, Gretarsson, Gretarsson, Griffith, Griffiths, Griggs, Grignani, Grimaldi, Grimm, Grote, Grunewald, Gruning, Guerra, Guidi, Guimaraes, Guix{\'e}, Gulati, Guo, Guo, Gupta, Gupta, Gupta, Gustafson, Gustafson, Guzman, Ha, Haegel, Hagiwara, Haino, Halim, Hall, Hamilton, Hammond, Han, Haney, Hanks, Hanna, Hannam, Hannuksela,
  Hansen, Hansen, Hanson, Harder, Hardwick, Haris, Harms, Harry, Harry, Hartwig, Hasegawa, Haskell, Hasskew, Haster, Hattori, Haughian, Hayakawa, Hayama, Hayes, Healy, Heidmann, Heidt, Heintze, Heinze, Heinzel, Heitmann, Hellman, Hello, {Helmling-Cornell}, Hemming, Hendry, Heng, Hennes, Hennig, Hennig, Hernandez, Vivanco, Heurs, Hild, Hill, Himemoto, Hines, Hiranuma, Hirata, Hirose, Ho, Hochheim, Hofman, Hohmann, Holcomb, Holland, Hollows, Holmes, Holt, Holz, Hong, Hopkins, Hough, Hourihane, Howell, Hoy, Hoyland, Hreibi, Hsieh, Hsu, Huang, Huang, Huang, Huang, Huang, Huang, H{\"u}bner, Huddart, Hughey, Hui, Hui, Husa, Huttner, Huxford, {Huynh-Dinh}, Ide, Idzkowski, Iess, Ikenoue, Imam, Inayoshi, Ingram, Inoue, Ioka, Isi, Isleif, Ito, Itoh, Iyer, Izumi, JaberianHamedan, Jacqmin, Jadhav, Jadhav, James, Jan, Jani, Janquart, Janssens, Janthalur, Jaranowski, Jariwala, Jaume, Jenkins, Jenner, Jeon, Jeunon, Jia, Jin, Johns, Jones, Jones, Jones, Jones, Jones, Jonker, Ju, Jung, Jung, Junker, Juste, Kaihotsu, Kajita,
  Kakizaki, Kalaghatgi, Kalogera, Kamai, Kamiizumi, Kanda, Kandhasamy, Kang, Kanner, Kao, Kapadia, Kapasi, Karat, Karathanasis, Karki, Kashyap, Kasprzack, Kastaun, Katsanevas, Katsavounidis, Katzman, Kaur, Kawabe, Kawaguchi, Kawai, Kawasaki, K{\'e}f{\'e}lian, Keitel, Key, Khadka, Khalili, Khan, Khazanov, Khetan, Khursheed, Kijbunchoo, Kim, Kim, Kim, Kim, Kim, Kim, Kimball, Kimura, {Kinley-Hanlon}, Kirchhoff, Kissel, Kita, Kitazawa, Kleybolte, Klimenko, Knee, Knowles, Knyazev, Koch, Koekoek, Kojima, Kokeyama, Koley, Kolitsidou, Kolstein, Komori, Kondrashov, Kong, Kontos, Koper, Korobko, Kotake, Kovalam, Kozak, Kozakai, Kozu, Kringel, Krishnendu, Kr{\'o}lak, Kuehn, Kuei, Kuijer, Kumar, Kumar, Kumar, Kumar, Kume, Kuns, Kuo, Kuo, Kuromiya, Kuroyanagi, Kusayanagi, Kuwahara, Kwak, Lagabbe, Laghi, Lalande, Lam, Lamberts, Landry, Lane, Lang, Lange, Lantz, Rosa, {Lartaux-Vollard}, Lasky, Laxen, Lazzarini, Lazzaro, Leaci, Leavey, Lecoeuche, Lee, Lee, Lee, Lee, Lee, Lee, Lehmann, Lema{\^i}tre, Leonardi, Leroy, Letendre,
  Levesque, Levin, Leviton, Leyde, Li, Li, Li, Li, Li, Li, Lin, Lin, Lin, Lin, Lin, Linde, Linker, Linley, Littenberg, Liu, Liu, Liu, Liu, Llamas, {Llorens-Monteagudo}, Lo, Lockwood, London, Longo, Lopez, Portilla, Lorenzini, Loriette, Lormand, Losurdo, Lott, Lough, Lousto, Lovelace, Lucaccioni, L{\"u}ck, Lumaca, Lundgren, Luo, Lynam, Macas, MacInnis, Macleod, MacMillan, Macquet, Hernandez, Magazz{\`u}, Magee, Maggiore, Magnozzi, Mahesh, Majorana, Makarem, Maksimovic, Maliakal, Malik, Man, Mandic, Mangano, Mango, Mansell, Manske, Mantovani, Mapelli, Marchesoni, Marchio, Marion, Mark, M{\'a}rka, M{\'a}rka, Markakis, Markosyan, Markowitz, Maros, Marquina, Marsat, Martelli, Martin, Martin, Martinez, Martinez, Martinez, Martinovic, Martynov, Marx, Masalehdan, Mason, Massera, Masserot, Massinger, {Masso-Reid}, Mastrogiovanni, Matas, {Mateu-Lucena}, Matichard, Matiushechkina, Mavalvala, McCann, McCarthy, McClelland, McClincy, McCormick, McCuller, McGhee, McGuire, McIsaac, McIver, McRae, McWilliams, Meacher, Mehmet,
  Mehta, Meijer, Melatos, Melchor, Mendell, {Menendez-Vazquez}, Menoni, Mercer, Mereni, Merfeld, Merilh, Merritt, Merzougui, Meshkov, Messenger, Messick, Meyers, Meylahn, Mhaske, Miani, Miao, Michaloliakos, Michel, Michimura, Middleton, Milano, Miller, Miller, Miller, Millhouse, Mills, Milotti, Minazzoli, Minenkov, Mio, Mir, {Miravet-Ten{\'e}s}, Mishra, Mishra, Mistry, Mitra, Mitrofanov, Mitselmakher, Mittleman, Miyakawa, Miyamoto, Miyazaki, Miyo, Miyoki, Mo, Modafferi, Moguel, Mogushi, Mohapatra, Mohite, Molina, {Molina-Ruiz}, Mondin, Montani, Moore, Moragues, Moraru, Morawski, More, Moreno, Moreno, Mori, Morisaki, Moriwaki, Mours, {Mow-Lowry}, Mozzon, Muciaccia, Mukherjee, Mukherjee, Mukherjee, Mukherjee, Mukherjee, Mukund, Mullavey, Munch, Mu{\~n}iz, Murray, Musenich, Muusse, Nadji, Nagano, Nagano, Nagar, Nakamura, Nakano, Nakano, Nakashima, Nakayama, Napolano, Nardecchia, Narikawa, Naticchioni, Nayak, Nayak, Negishi, Neil, Neilson, Nelemans, Nelson, Nery, Neubauer, Neunzert, Ng, Ng, Nguyen, Nguyen,
  Nguyen, Quynh, Ni, Nichols, Nishizawa, Nissanke, Nitoglia, Nocera, Norman, North, Nozaki, Nuttall, Oberling, O'Brien, Obuchi, O'Dell, Oelker, Ogaki, Oganesyan, Oh, Oh, Oh, Ohashi, Ohishi, Ohkawa, Ohme, Ohta, Okada, Okutani, Okutomi, Olivetto, Oohara, Ooi, Oram, O'Reilly, Ormiston, Ormsby, Ortega, O'Shaughnessy, O'Shea, Oshino, Ossokine, Osthelder, Otabe, Ottaway, Overmier, Pace, Pagano, Page, Pagliaroli, Pai, Pai, Palamos, Palashov, Palomba, Pan, Pan, Panda, Pang, Pang, Pankow, Pannarale, Pant, Panther, Paoletti, Paoli, Paolone, Parisi, Park, Park, Parker, Pascucci, Pasqualetti, Passaquieti, Passuello, Patel, Pathak, Patricelli, Patron, Patrone, Paul, Payne, Pedraza, Pegoraro, Pele, Arellano, Penn, Perego, Pereira, Pereira, Perez, P{\'e}rigois, Perkins, Perreca, Perri{\`e}s, Petermann, Petterson, Pfeiffer, Pham, Phukon, Piccinni, Pichot, Piendibene, Piergiovanni, Pierini, Pierro, Pillant, Pillas, Pilo, Pinard, Pinto, Pinto, Piotrzkowski, Pirello, Pitkin, Placidi, Planas, Plastino, Pluchar, Poggiani, Polini,
  Pong, Ponrathnam, Popolizio, Porter, Poulton, Powell, Pracchia, Pradier, Prajapati, Prasai, Prasanna, Pratten, Principe, Prodi, Prokhorov, Prosposito, Prudenzi, Puecher, Punturo, Puosi, Puppo, P{\"u}rrer, Qi, Quetschke, {Quitzow-James}, Raab, Raaijmakers, Radkins, Radulesco, Raffai, Rail, Raja, Rajan, Ramirez, Ramirez, {Ramos-Buades}, Rana, Rapagnani, Rapol, Ray, Raymond, Raza, Razzano, Read, Rees, Regimbau, Rei, Reid, Reid, Reitze, Relton, Renzini, Rettegno, Rezac, Ricci, Richards, Richardson, Richardson, Riemenschneider, Riles, Rinaldi, Rink, Rizzo, Robertson, Robie, Robinet, Rocchi, Rodriguez, Rolland, Rollins, Romanelli, Romano, Romel, {Romero-Rodr{\'i}guez}, {Romero-Shaw}, Romie, Ronchini, Rosa, Rose, Rosi{\'n}ska, Ross, Rowan, Rowlinson, Roy, Roy, Roy, Rozza, Ruggi, Ryan, Sachdev, Sadecki, Sadiq, Sago, Saito, Saito, Sakai, Sakai, Sakellariadou, Sakuno, Salafia, Salconi, Saleem, Salemi, Samajdar, Sanchez, Sanchez, Sanchez, {Sanchis-Gual}, Sanders, Sanuy, Saravanan, Sarin, Sassolas, Satari, Sato, Sato,
  Sauter, Savage, Sawada, Sawant, Sawant, Sayah, Schaetzl, Scheel, Scheuer, Schiworski, Schmidt, Schmidt, Schnabel, Schneewind, Schofield, Sch{\"o}nbeck, Schulte, Schutz, Schwartz, Scott, Scott, {Seglar-Arroyo}, Sekiguchi, Sekiguchi, Sellers, Sengupta, Sentenac, Seo, Sequino, Sergeev, Setyawati, Shaffer, Shahriar, Shams, Shao, Sharma, Sharma, Shawhan, Shcheblanov, Shibagaki, Shikauchi, Shimizu, Shimoda, Shimode, Shinkai, Shishido, Shoda, Shoemaker, Shoemaker, ShyamSundar, Sieniawska, Sigg, Singer, Singh, Singh, Singha, Sintes, Sipala, Skliris, Slagmolen, {Slaven-Blair}, Smetana, Smith, Smith, Soldateschi, Somala, Somiya, Son, Soni, Soni, Sordini, Sorrentino, Sorrentino, Sotani, Soulard, Souradeep, Sowell, Spagnuolo, Spencer, Spera, Srinivasan, Srivastava, Srivastava, Staats, Stachie, Steer, Steinlechner, Steinlechner, Stops, Stover, Strain, Strang, Stratta, Strunk, Sturani, Stuver, Sudhagar, Sudhir, Sugimoto, Suh, Summerscales, Sun, Sun, Sunil, Sur, Suresh, Sutton, Suzuki, Suzuki, Swinkels, Szczepa{\'n}czyk,
  Szewczyk, Tacca, Tagoshi, Tait, Takahashi, Takahashi, Takamori, Takano, Takeda, Takeda, Talbot, Talbot, Tanaka, Tanaka, Tanaka, Tanaka, Tanaka, Tanasijczuk, Tanioka, Tanner, Tao, Tao, San~Mart{\'i}n, Taranto, Tasson, Telada, Tenorio, Terhune, Terkowski, Thirugnanasambandam, Thomas, Thomas, Thompson, Thondapu, Thorne, Thrane, Tiwari, Tiwari, Tiwari, Toivonen, Toland, Tolley, Tomaru, Tomigami, Tomura, Tonelli, {Torres-Forn{\'e}}, Torrie, E~Melo, T{\"o}yr{\"a}, Trapananti, Travasso, Traylor, Trevor, Tringali, Tripathee, Troiano, Trovato, Trozzo, Trudeau, Tsai, Tsai, Tsang, Tsang, Tsao, Tse, Tso, Tsubono, Tsuchida, Tsukada, Tsuna, Tsutsui, Tsuzuki, Turbang, Turconi, Tuyenbayev, Ubhi, Uchikata, Uchiyama, Udall, Ueda, Uehara, Ueno, Ueshima, Unnikrishnan, Uraguchi, Urban, Ushiba, Utina, Vahlbruch, Vajente, Vajpeyi, Valdes, Valentini, Valsan, Van~Bakel, Van~Beuzekom, Van Den~Brand, Van Den~Broeck, {Vander-Hyde}, Van Der~Schaaf, Van~Heijningen, Vanosky, Van~Putten, Van~Remortel, Vardaro, Vargas, Varma, Vas{\'u}th,
  Vecchio, Vedovato, Veitch, Veitch, Venneberg, Venugopalan, Verkindt, Verma, Verma, Veske, Vetrano, Vicer{\'e}, Vidyant, Viets, Vijaykumar, {Villa-Ortega}, Vinet, Virtuoso, Vitale, Vo, Vocca, Reis, Wrangel, Vorvick, Vyatchanin, Wade, Wade, Wagner, Walet, Walker, Wallace, Wallace, Walsh, Wang, Wang, Wang, Ward, Warner, Was, Washimi, Washington, Watchi, Weaver, Webster, Weinert, Weinstein, Weiss, Weller, Wellmann, Wen, We{\ss}els, Wette, Whelan, White, Whiting, Whittle, Wilken, Williams, Williams, Williamson, Willis, Willke, Wilson, Winkler, Wipf, Wlodarczyk, Woan, Woehler, Wofford, Wong, Wu, Wu, Wu, Wu, Wysocki, Xiao, Xu, Yamada, Yamamoto, Yamamoto, Yamamoto, Yamamoto, Yamashita, Yamazaki, Yang, Yang, Yang, Yang, Yang, Yap, Yeeles, Yelikar, Ying, Yokogawa, Yokoyama, Yokozawa, Yoo, Yoshioka, Yu, Yu, Yuzurihara, Zadro{\.z}ny, Zanolin, Zeidler, Zelenova, Zendri, Zevin, Zhan, Zhang, Zhang, Zhang, Zhang, Zhang, Zhao, Zhao, Zhao, Zhao, Zhou, Zhou, Zhu, Zhu, Zucker, Zweizig, Antonopoulou, Arzoumanian, Basu,
  Bogdanov, Cognard, Crowter, Enoto, Espinoza, Flynn, Fonseca, Good, Guillemot, Guillot, Harding, Keith, Kuiper, Lower, Lyne, McKee, Meyers, Ng, Palfreyman, Shannon, Shaw, Stairs, Stappers, Tan, Theureau, \& Weltevrede}]{abbott2022Narrowband}
Abbott, R., Abbott, T.~D., Acernese, F., {et~al.} 2022{\natexlab{a}}, The Astrophysical Journal, 932, 133, \dodoi{10.3847/1538-4357/ac6ad0}

\bibitem[{Abbott {et~al.}(2022{\natexlab{b}})Abbott, Abe, Acernese, Ackley, Adhikari, Adhikari, Adkins, Adya, Affeldt, Agarwal, Agathos, Agatsuma, Aggarwal, Aguiar, Aiello, Ain, Ajith, Akutsu, Albanesi, Alfaidi, Allocca, Altin, Amato, Anand, Anand, Ananyeva, Anderson, Anderson, Ando, Andrade, Andres, {Andr{\'e}s-Carcasona}, Andri{\'c}, Angelova, Ansoldi, Antelis, Antier, Apostolatos, Appavuravther, Appert, Apple, Arai, Araya, Araya, Areeda, Ar{\`e}ne, Aritomi, Arnaud, Arogeti, Aronson, Asada, Asali, Ashton, Aso, Assiduo, Assis De Souza~Melo, Aston, Astone, Aubin, AultONeal, Austin, Babak, Badaracco, Bader, Badger, Bae, Bae, Baer, Bagnasco, Bai, Baird, Bajpai, Baka, Ball, Ballardin, Ballmer, Balsamo, Baltus, Banagiri, Banerjee, Bankar, Barayoga, Barbieri, Barish, Barker, Barneo, Barone, Barr, Barsotti, Barsuglia, Barta, Bartlett, Barton, Bartos, Basak, Bassiri, Basti, Bawaj, Bayley, Bazzan, Becher, B{\'e}csy, Bedakihale, Beirnaert, Bejger, Belahcene, Benedetto, Beniwal, Benjamin, Bennett, Bentley, BenYaala,
  Bera, Berbel, Bergamin, Berger, Bernuzzi, Bersanetti, Bertolini, Betzwieser, Beveridge, Bhandare, Bhandari, Bhardwaj, Bhatt, Bhattacharjee, Bhaumik, Bianchi, Bilenko, Billingsley, Bini, Birney, Birnholtz, Biscans, Bischi, Biscoveanu, Bisht, Biswas, Bitossi, Bizouard, Blackburn, Blair, Blair, Blair, Bobba, Bode, Bo{\"e}r, Bogaert, Boldrini, Bolingbroke, Bonavena, Bondu, Bonilla, Bonnand, Booker, Boom, Bork, Boschi, Bose, Bose, Bossilkov, Boudart, Bouffanais, Bozzi, Bradaschia, Brady, Bramley, Branch, Branchesi, Brau, Breschi, Briant, Briggs, Brillet, Brinkmann, Brockill, Brooks, Brooks, Brown, Brunett, Bruno, Bruntz, Bryant, Bucci, Bulik, Bulten, Buonanno, Burtnyk, Buscicchio, Buskulic, Buy, Byer, Cabourn~Davies, Cabras, Cabrita, Cadonati, Caesar, Cagnoli, Cahillane, Calder{\'o}n~Bustillo, Callaghan, Callister, Calloni, Cameron, Camp, Canepa, Canevarolo, Cannavacciuolo, Cannon, Cao, Cao, Capocasa, Capote, Carapella, Carbognani, Carlassara, Carlin, Carney, Carpinelli, Carrillo, Carullo, Carver,
  Casanueva~Diaz, Casentini, Castaldi, Caudill, Cavagli{\`a}, Cavalier, Cavalieri, Cella, {Cerd{\'a}-Dur{\'a}n}, Cesarini, Chaibi, Chalathadka~Subrahmanya, Champion, Chan, Chan, Chan, Chan, Chan, Chandra, Chang, Chanial, Chao, {Chapman-Bird}, Charlton, Chase, {Chassande-Mottin}, Chatterjee, Chatterjee, Chatterjee, Chaturvedi, Chaty, Chen, Chen, Chen, Chen, Chen, Chen, Chen, Chen, Chen, Cheng, Cheong, Cheung, Chia, Chiadini, Chiang, Chiarini, Chierici, Chincarini, Chiofalo, Chiummo, Choudhary, Choudhary, Christensen, Chu, Chu, Chua, Chung, Ciani, Ciecielag, Cie{\'s}lar, Cifaldi, Ciobanu, Ciolfi, Cipriano, Clara, Clark, Clearwater, Clesse, Cleva, Coccia, Codazzo, Cohadon, Cohen, Colleoni, Collette, Colombo, Colpi, Compton, Constancio, Conti, Cooper, Corban, Corbitt, {Cordero-Carri{\'o}n}, Corezzi, Corley, Cornish, Corre, Corsi, Cortese, Costa, Cotesta, Cottingham, Coughlin, Coulon, Countryman, Cousins, Couvares, Coward, Cowart, Coyne, Coyne, Creighton, Creighton, Criswell, Croquette, Crowder, Cudell, Cullen,
  Cumming, Cummings, Cunningham, Cuoco, Cury{\l}o, Dabadie, Dal~Canton, Dall'Osso, D{\'a}lya, Dana, D'Angelo, Danilishin, D'Antonio, Danzmann, {Darsow-Fromm}, Dasgupta, Datrier, Datta, Datta, Dattilo, Dave, Davier, Davis, Davis, Daw, Dean, DeBra, Deenadayalan, Degallaix, De~Laurentis, Del{\'e}glise, Del~Favero, De~Lillo, De~Lillo, Dell'Aquila, Del~Pozzo, DeMarchi, De~Matteis, D'Emilio, Demos, Dent, Depasse, De~Pietri, De~Rosa, De~Rossi, DeSalvo, De~Simone, Dhurandhar, D{\'i}az, Di~Cesare, Didio, Dietrich, Di~Fiore, Di~Fronzo, Di~Giorgio, Di~Giovanni, Di~Giovanni, Di~Girolamo, Di~Lieto, Di~Michele, Ding, Di~Pace, Di~Palma, Di~Renzo, Divakarla, Dmitriev, Doctor, Donahue, D'Onofrio, Donovan, Dooley, Doravari, Dorosh, Drago, Driggers, Drori, Ducoin, Dupej, Dupletsa, Durante, D'Urso, Duverne, Dwyer, Eassa, Easter, Ebersold, Eckhardt, Eddolls, Edelman, Edo, Edy, Effler, Eguchi, Eichholz, Eikenberry, Eisenmann, Eisenstein, Ejlli, Engelby, Enomoto, Errico, Essick, Estell{\'e}s, Estevez, Etienne, Etzel, Evans, Evans,
  Evstafyeva, Ewing, Fabrizi, Faedi, Fafone, Fair, Fairhurst, Fan, Farah, Farinon, Farr, Farr, {Fauchon-Jones}, Favaro, Favata, Fays, Fazio, Feicht, Fejer, Fenyvesi, Ferguson, {Fernandez-Galiana}, Ferrante, Ferreira, Fidecaro, Figura, Fiori, Fiori, Fishbach, Fisher, Fittipaldi, Fiumara, Flaminio, Floden, Fong, Font, Fornal, Forsyth, Franke, Frasca, Frasconi, Freed, Frei, Freise, Freitas, Frey, Fritschel, Frolov, Fronz{\'e}, Fujii, Fujikawa, Fujimoto, Fulda, Fyffe, Gabbard, Gadre, Gair, Gais, Galaudage, Gamba, Ganapathy, Ganguly, Gao, Gaonkar, Garaventa, Garc{\'i}a~N{\'u}{\~n}ez, {Garc{\'i}a-Quir{\'o}s}, Garufi, Gateley, Gayathri, Ge, Gemme, Gennai, George, Gerberding, Gergely, Gewecke, Ghonge, Ghosh, Ghosh, Ghosh, Ghosh, Ghosh, Giacomazzo, Giacoppo, Giaime, Giardina, Gibson, Gier, Giesler, Giri, Gissi, Gkaitatzis, Glanzer, Gleckl, Godwin, Goetz, Goetz, Gohlke, Golomb, Goncharov, Gonz{\'a}lez, Gosselin, Gouaty, Gould, Goyal, Grace, Grado, Graham, Granata, Granata, Grant, Gras, Grassia, Gray, Gray, Greco,
  Green, Green, Gretarsson, Gretarsson, Griffith, Griffiths, Griggs, Grignani, Grimaldi, Grimes, Grimm, Grote, Grunewald, Gruning, Gruson, Guerra, Guidi, Guimaraes, Guix{\'e}, Gulati, Gunny, Guo, Guo, Gupta, Gupta, Gupta, Gupta, Gupta, Gustafson, Guzman, Ha, Hadiputrawan, Haegel, Haino, Halim, Hall, Hamilton, Hammond, Han, Haney, Hanks, Hanna, Hannam, Hannuksela, Hansen, Hansen, Hanson, Harder, Haris, Harms, Harry, Harry, Hartwig, Hasegawa, Haskell, Haster, Hathaway, Hattori, Haughian, Hayakawa, Hayama, Hayes, Healy, Heidmann, Heidt, Heintze, Heinze, Heinzel, Heitmann, Hellman, Hello, {Helmling-Cornell}, Hemming, Hendry, Heng, Hennes, Hennig, Hennig, Henshaw, Hernandez, Hernandez~Vivanco, Heurs, Hewitt, Higginbotham, Hild, Hill, Himemoto, Hines, Hirata, Hirose, Ho, Hochheim, Hofman, Hohmann, Holcomb, Holland, Hollows, Holmes, Holt, Holz, Hong, Hough, Hourihane, Howell, Hoy, Hoyland, Hreibi, Hsieh, Hsieh, Hsiung, Hsu, Huang, Huang, Huang, Huang, Huang, Huang, H{\"u}bner, Huddart, Hughey, Hui, Hui, Husa,
  Huttner, Huxford, {Huynh-Dinh}, Ide, Idzkowski, Iess, Inayoshi, Inoue, Iosif, Isi, Isleif, Ito, Itoh, Iyer, JaberianHamedan, Jacqmin, Jacquet, Jadhav, Jadhav, Jain, James, Jan, Jani, Janquart, Janssens, Janthalur, Jaranowski, Jariwala, Jaume, Jenkins, Jenner, Jeon, Jia, Jiang, Jin, Johns, Johnston, Jones, Jones, Jones, Jones, Joshi, Ju, Jue, Jung, Jung, Junker, Juste, Kaihotsu, Kajita, Kakizaki, Kalaghatgi, Kalogera, Kamai, Kamiizumi, Kanda, Kandhasamy, Kang, Kanner, Kao, Kapadia, Kapasi, Karathanasis, Karki, Kashyap, Kasprzack, Kastaun, Kato, Katsanevas, Katsavounidis, Katzman, Kaur, Kawabe, Kawaguchi, K{\'e}f{\'e}lian, Keitel, Key, Khadka, Khalili, Khan, Khanam, Khazanov, Khetan, Khursheed, Kijbunchoo, Kim, Kim, Kim, Kim, Kim, Kim, Kim, Kimball, Kimura, {Kinley-Hanlon}, Kirchhoff, Kissel, Klimenko, Klinger, Knee, Knowles, Knust, Knyazev, Kobayashi, Koch, Koekoek, Kohri, Kokeyama, Koley, Kolitsidou, Kolstein, Komori, Kondrashov, Kong, Kontos, Koper, Korobko, Kovalam, Koyama, Kozak, Kozakai, Kringel,
  Krishnendu, Kr{\'o}lak, Kuehn, Kuei, Kuijer, Kulkarni, Kumar, Kumar, Kumar, Kumar, Kume, Kuns, Kuromiya, Kuroyanagi, Kwak, Lacaille, Lagabbe, Laghi, Lalande, Lalleman, Lam, Lamberts, Landry, Lane, Lang, Lange, Lantz, La~Rosa, {Lartaux-Vollard}, Lasky, Laxen, Lazzarini, Lazzaro, Leaci, Leavey, LeBohec, Lecoeuche, Lee, Lee, Lee, Lee, Lee, Legred, Lehmann, Lema{\^i}tre, Lenti, Leonardi, Leonova, Leroy, Letendre, Levesque, Levin, Leviton, Leyde, Li, Li, Li, Li, Li, Li, Li, Lin, Lin, Lin, Lin, Lin, Lin, Linde, Linker, Linley, Littenberg, Liu, Liu, Liu, Liu, Llamas, Lo, Lo, London, Longo, Lopez, Lopez~Portilla, Lorenzini, Loriette, Lormand, Losurdo, Lott, Lough, Lousto, Lovelace, Lucaccioni, L{\"u}ck, Lumaca, Lundgren, Luo, Lynam, Ma'arif, Macas, Machtinger, MacInnis, Macleod, MacMillan, Macquet, Maga{\~n}a~Hernandez, Magazz{\`u}, Magee, Maggiore, Magnozzi, Mahesh, Majorana, Maksimovic, Maliakal, Malik, Man, Mandic, Mangano, Mansell, Manske, Mantovani, Mapelli, Marchesoni, Mar{\'i}n~Pina, Marion, Mark, M{\'a}rka,
  M{\'a}rka, Markakis, Markosyan, Markowitz, Maros, Marquina, Marsat, Martelli, Martin, Martin, Martinez, Martinez, Martinez, Martinovic, Martynov, Marx, Masalehdan, Mason, Massera, Masserot, {Masso-Reid}, Mastrogiovanni, Matas, {Mateu-Lucena}, Matichard, Matiushechkina, Mavalvala, McCann, McCarthy, McClelland, McClincy, McCormick, McCuller, McGhee, McGuire, McIsaac, McIver, McRae, McWilliams, Meacher, Mehmet, Mehta, Meijer, Melatos, Melchor, Mendell, {Menendez-Vazquez}, Menoni, Mercer, Mereni, Merfeld, Merilh, Merritt, Merzougui, Meshkov, Messenger, Messick, Meyers, Meylahn, Mhaske, Miani, Miao, Michaloliakos, Michel, Michimura, Middleton, Mihaylov, Milano, Miller, Miller, Miller, Millhouse, Mills, Milotti, Minenkov, Mio, Mir, {Miravet-Ten{\'e}s}, Mishkin, Mishra, Mishra, Mistry, Mitra, Mitrofanov, Mitselmakher, Mittleman, Miyakawa, Miyo, Miyoki, Mo, Modafferi, Moguel, Mogushi, Mohapatra, Mohite, Molina, {Molina-Ruiz}, Mondin, Montani, Moore, Moragues, Moraru, Morawski, More, Moreno, Moreno, Mori, Morisaki,
  Morisue, Moriwaki, Mours, {Mow-Lowry}, Mozzon, Muciaccia, Mukherjee, Mukherjee, Mukherjee, Mukherjee, Mukherjee, Mukund, Mullavey, Munch, Mu{\~n}iz, Murray, Musenich, Muusse, Nadji, Nagano, Nagar, Nakamura, Nakano, Nakano, Nakayama, Napolano, Nardecchia, Narikawa, Narola, Naticchioni, Nayak, Nayak, Neil, Neilson, Nelson, Nelson, Nery, Neubauer, Neunzert, Ng, Ng, Nguyen, Nguyen, Nguyen, Quynh, Ni, Ni, Nichols, Nishimoto, Nishizawa, Nissanke, Nitoglia, Nocera, Norman, North, Nozaki, Nurbek, Nuttall, Obayashi, Oberling, O'Brien, O'Dell, Oelker, Ogaki, Oganesyan, Oh, Oh, Oh, Ohashi, Ohashi, Ohkawa, Ohme, Ohta, Okada, Okutani, Olivetto, Oohara, Oram, O'Reilly, Ormiston, Ormsby, O'Shaughnessy, O'Shea, Oshino, Ossokine, Osthelder, Otabe, Ottaway, Overmier, Pace, Pagano, Pagano, Page, Pagliaroli, Pai, Pai, Pal, Palamos, Palashov, Palomba, Pan, Pan, Panda, Pang, Pankow, Pannarale, Pant, Panther, Paoletti, Paoli, Paolone, Pappas, Parisi, Park, Park, Parker, Pascucci, Pasqualetti, Passaquieti, Passuello, Patel,
  Pathak, Patricelli, Patron, Paul, Payne, Pedraza, Pedurand, Pegoraro, Pele, Pe{\~n}a~Arellano, Penano, Penn, Perego, Pereira, Pereira, Perez, P{\'e}rigois, Perkins, Perreca, Perri{\`e}s, Pesios, Petermann, Petterson, Pfeiffer, Pham, Pham, Phukon, Phurailatpam, Piccinni, Pichot, Piendibene, Piergiovanni, Pierini, Pierro, Pillant, Pillas, Pilo, Pinard, {Pineda-Bosque}, Pinto, Pinto, Piotrzkowski, Piotrzkowski, Pirello, Pisarski, Pitkin, Placidi, Placidi, Planas, Plastino, Pluchar, Poggiani, Polini, Pong, Ponrathnam, Porter, Poulton, Poverman, Powell, Pracchia, Pradier, Prajapati, Prasai, Prasanna, Pratten, Principe, Prodi, Prokhorov, Prosposito, Prudenzi, Puecher, Punturo, Puosi, Puppo, P{\"u}rrer, Qi, Quartey, Quetschke, Quinonez, {Quitzow-James}, Raab, Raaijmakers, Radkins, Radulesco, Raffai, Rail, Raja, Rajan, Ramirez, Ramirez, {Ramos-Buades}, Rana, Rapagnani, Ray, Raymond, Raza, Razzano, Read, Rees, Regimbau, Rei, Reid, Reid, Reitze, Relton, Renzini, Rettegno, Revenu, Reza, Rezac, Ricci, Richards,
  Richardson, Richardson, Riemenschneider, Riles, Rinaldi, Rink, Robertson, Robie, Robinet, Rocchi, Rodriguez, Rolland, Rollins, Romanelli, Romano, Romel, Romero, {Romero-Shaw}, Romie, Ronchini, Rosa, Rose, Rosi{\'n}ska, Ross, Rowan, Rowlinson, Roy, Roy, Roy, Rozza, Ruggi, {Ruiz-Rocha}, Ryan, Sachdev, Sadecki, Sadiq, Saha, Saito, Sakai, Sakellariadou, Sakon, Salafia, {Salces-Carcoba}, Salconi, Saleem, Salemi, Samajdar, Sanchez, Sanchez, Sanchez, {Sanchis-Gual}, Sanders, Sanuy, Saravanan, Sarin, Sassolas, Satari, Sauter, Savage, Savant, Sawada, Sawant, Sayah, Schaetzl, Scheel, Scheuer, Schiworski, Schmidt, Schmidt, Schnabel, Schneewind, Schofield, Sch{\"o}nbeck, Schulte, Schutz, Schwartz, Scott, Scott, {Seglar-Arroyo}, Sekiguchi, Sellers, Sengupta, Sentenac, Seo, Sequino, Sergeev, Setyawati, Shaffer, Shahriar, Shaikh, Shams, Shao, Sharma, Sharma, Shawhan, Shcheblanov, Sheela, Shikano, Shikauchi, Shimizu, Shimode, Shinkai, Shishido, Shoda, Shoemaker, Shoemaker, ShyamSundar, Sieniawska, Sigg, Silenzi, Singer,
  Singh, Singh, Singh, Singha, Sintes, Sipala, Skliris, Slagmolen, {Slaven-Blair}, Smetana, Smith, Smith, Smith, Soldateschi, Somala, Somiya, Song, Soni, Soni, Sordini, Sorrentino, Sorrentino, Soulard, Souradeep, Sowell, Spagnuolo, Spencer, Spera, Spinicelli, Srivastava, Srivastava, Staats, Stachie, Stachurski, Steer, Steinlechner, Steinlechner, Stergioulas, Stops, Stover, Strain, Strang, Stratta, Strong, Strunk, Sturani, Stuver, Suchenek, Sudhagar, Sudhir, Sugimoto, Suh, Sullivan, Summerscales, Sun, Sunil, Sur, Suresh, Sutton, Suzuki, Suzuki, Suzuki, Swinkels, Szczepa{\'n}czyk, Szewczyk, Tacca, Tagoshi, Tait, Takahashi, Takahashi, Takano, Takeda, Takeda, Talbot, Talbot, Tanaka, Tanaka, Tanaka, Tanasijczuk, Tanioka, Tanner, Tao, Tao, Tapia, Tapia San~Mart{\'i}n, Taranto, Taruya, Tasson, Tenorio, Terhune, Terkowski, Thirugnanasambandam, Thomas, Thomas, Thompson, Thompson, Thondapu, Thorne, Thrane, Tiwari, Tiwari, Tiwari, Toivonen, Tolley, Tomaru, Tomura, Tonelli, Tornasi, {Torres-Forn{\'e}}, Torrie, Tosta
  E~Melo, T{\"o}yr{\"a}, Trapananti, Travasso, Traylor, Trevor, Tringali, Tripathee, Troiano, Trovato, Trozzo, Trudeau, Tsai, Tsang, Tsang, Tsao, Tse, Tso, Tsuchida, Tsukada, Tsuna, Tsutsui, Turbang, Turconi, Tuyenbayev, Ubhi, Uchikata, Uchiyama, Udall, Ueda, Uehara, Ueno, Ueshima, Unnikrishnan, Urban, Ushiba, Utina, Vajente, Vajpeyi, Valdes, Valentini, Valsan, Van~Bakel, Van~Beuzekom, Van~Dael, Van Den~Brand, Van Den~Broeck, {Vander-Hyde}, Van~Haevermaet, Van~Heijningen, Van~Putten, Van~Remortel, Vardaro, Vargas, Varma, Vas{\'u}th, Vecchio, Vedovato, Veitch, Veitch, Venneberg, Venugopalan, Verkindt, Verma, Verma, Vermeulen, Veske, Vetrano, Vicer{\'e}, Vidyant, Viets, Vijaykumar, {Villa-Ortega}, Vinet, Virtuoso, Vitale, Vocca, Von~Reis, Von~Wrangel, Vorvick, Vyatchanin, Wade, Wade, Wagner, Walet, Walker, Wallace, Wallace, Wang, Wang, Wang, Ward, Warner, Was, Washimi, Washington, Watchi, Weaver, Weaving, Webster, Weinert, Weinstein, Weiss, Weller, Weller, Wellmann, Wen, We{\ss}els, Wette, Whelan, White,
  Whiting, Whittle, Wilken, Williams, Williams, Williamson, Willis, Willke, Wilson, Wipf, Wlodarczyk, Woan, Woehler, Wofford, Wong, Wong, Wright, Wu, Wu, Wu, Wysocki, Xiao, Yamada, Yamamoto, Yamamoto, Yamamoto, Yamashita, Yamazaki, Yang, Yang, Yang, Yang, Yang, Yang, Yap, Yeeles, Yeh, Yelikar, Ying, Yokoyama, Yokozawa, Yoo, Yoshioka, Yu, Yu, Yuzurihara, Zadro{\.z}ny, Zanolin, Zeidler, Zelenova, Zendri, Zevin, Zhan, Zhang, Zhang, Zhang, Zhang, Zhang, Zhang, Zhao, Zhao, Zhao, Zhao, Zhou, Zhou, Zhu, Zhu, Zucker, Zweizig, \& {LIGO Scientific Collaboration, Virgo Collaboration, and KAGRA Collaboration}}]{abbott2022Allsky}
Abbott, R., Abe, H., Acernese, F., {et~al.} 2022{\natexlab{b}}, Physical Review D, 106, 102008, \dodoi{10.1103/PhysRevD.106.102008}

\bibitem[{Bonazzola \& Gourgoulhon(1996)}]{bonazzola1996Gravitational}
Bonazzola, S., \& Gourgoulhon, E. 1996, Astronomy and Astrophysics, 312, 675, \dodoi{10.48550/arXiv.astro-ph/9602107}

\bibitem[{Choudhuri \& Konar(2002)}]{choudhuri2002Diamagnetic}
Choudhuri, A., \& Konar, S. 2002, Monthly Notices of the Royal Astronomical Society, 332, 933, \dodoi{10.1046/j.1365-8711.2002.05362.x}

\bibitem[{Contopoulos {et~al.}(2023)Contopoulos, Kazanas, \& Papadopoulos}]{contopoulos2023Gravitational}
Contopoulos, I., Kazanas, D., \& Papadopoulos, D.~B. 2023, Monthly Notices of the Royal Astronomical Society, 527, 11198, \dodoi{10.1093/mnras/stad3913}

\bibitem[{Fujisawa {et~al.}(2022)Fujisawa, Kisaka, \& Kojima}]{fujisawa2022Magnetically}
Fujisawa, K., Kisaka, S., \& Kojima, Y. 2022, Monthly Notices of the Royal Astronomical Society, 516, 5196, \dodoi{10.1093/mnras/stac2585}

\bibitem[{Giliberti \& Cambiotti(2022)}]{giliberti2022Starquakes}
Giliberti, E., \& Cambiotti, G. 2022, Monthly Notices of the Royal Astronomical Society, 511, 3365, \dodoi{10.1093/mnras/stac245}

\bibitem[{Giliberti {et~al.}(2019)Giliberti, Cambiotti, Antonelli, \& Pizzochero}]{giliberti2019Modelling}
Giliberti, E., Cambiotti, G., Antonelli, M., \& Pizzochero, P.~M. 2019, Monthly Notices of the Royal Astronomical Society, stz3099, \dodoi{10.1093/mnras/stz3099}

\bibitem[{Gittins \& Andersson(2021)}]{gittins2021Modelling}
Gittins, F., \& Andersson, N. 2021, Monthly Notices of the Royal Astronomical Society, 507, 116, \dodoi{10.1093/mnras/stab2048}

\bibitem[{Gittins {et~al.}(2020)Gittins, Andersson, \& Jones}]{gittins2020Modelling}
Gittins, F., Andersson, N., \& Jones, D.~I. 2020, Monthly Notices of the Royal Astronomical Society, 500, 5570, \dodoi{10.1093/mnras/staa3635}

\bibitem[{Hacyan(2017)}]{hacyan2017Gravitational}
Hacyan, S. 2017, Revista Mexicana de F{\'i}sica, 3

\bibitem[{Haskell {et~al.}(2008)Haskell, Samuelsson, Glampedakis, \& Andersson}]{haskell2008Modelling}
Haskell, B., Samuelsson, L., Glampedakis, K., \& Andersson, N. 2008, Monthly Notices of the Royal Astronomical Society, 385, 531, \dodoi{10.1111/j.1365-2966.2008.12861.x}

\bibitem[{Kerin \& Melatos(2022)}]{kerin2022Mountain}
Kerin, A.~D., \& Melatos, A. 2022, Monthly Notices of the Royal Astronomical Society, 514, 1628, \dodoi{10.1093/mnras/stac1351}

\bibitem[{Maggiore(2007)}]{maggiore2007Gravitational}
Maggiore, M. 2007, Gravitational {{Waves}}: {{Volume}} 1: {{Theory}} and {{Experiments}}, 1st edn. (Oxford University PressOxford), \dodoi{10.1093/acprof:oso/9780198570745.001.0001}

\bibitem[{Melatos \& Payne(2005)}]{melatos2005Gravitational}
Melatos, A., \& Payne, D. J.~B. 2005, The Astrophysical Journal, 623, 1044, \dodoi{10.1086/428600}

\bibitem[{Melatos \& Phinney(2001)}]{melatos2001Hydromagnetic}
Melatos, A., \& Phinney, E.~S. 2001, Publications of the Astronomical Society of Australia, 18, 421, \dodoi{10.1071/AS01056}

\bibitem[{Mukherjee(2017)}]{mukherjee2017Revisiting}
Mukherjee, D. 2017, Journal of Astrophysics and Astronomy, 38, 48, \dodoi{10.1007/s12036-017-9465-6}

\bibitem[{Mukherjee \& Bhattacharya(2012)}]{mukherjee2012Phasedependent}
Mukherjee, D., \& Bhattacharya, D. 2012, Monthly Notices of the Royal Astronomical Society, 420, 720, \dodoi{10.1111/j.1365-2966.2011.20085.x}

\bibitem[{Mukherjee {et~al.}(2013)Mukherjee, Bhattacharya, \& Mignone}]{mukherjee2013MHD}
Mukherjee, D., Bhattacharya, D., \& Mignone, A. 2013, {{MHD}} Instabilities in Accretion Mounds on Neutron Star Binaries,  arXiv, \dodoi{10.48550/ARXIV.1304.7262}

\bibitem[{Nazari \& Roshan(2020)}]{nazari2020Gravitational}
Nazari, E., \& Roshan, M. 2020, Monthly Notices of the Royal Astronomical Society, 498, 110, \dodoi{10.1093/mnras/staa2322}

\bibitem[{Passamonti \& Lander(2014)}]{passamonti2014Quasiperiodic}
Passamonti, A., \& Lander, S.~K. 2014, Monthly Notices of the Royal Astronomical Society, 438, 156, \dodoi{10.1093/mnras/stt2134}

\bibitem[{Payne \& Melatos(2004)}]{pm04}
Payne, D. J.~B., \& Melatos, A. 2004, Monthly Notices of the Royal Astronomical Society, 351, 569, \dodoi{10.1111/j.1365-2966.2004.07798.x}

\bibitem[{Payne \& Melatos(2007)}]{payne2007Burial}
---. 2007, Monthly Notices of the Royal Astronomical Society, 376, 609, \dodoi{10.1111/j.1365-2966.2007.11451.x}

\bibitem[{Priymak {et~al.}(2011)Priymak, Melatos, \& Payne}]{priymak2011Quadrupole}
Priymak, M., Melatos, A., \& Payne, D. J.~B. 2011, Monthly Notices of the Royal Astronomical Society, 417, 2696, \dodoi{10.1111/j.1365-2966.2011.19431.x}

\bibitem[{Riles(2023)}]{riles2023Searches}
Riles, K. 2023, Living Reviews in Relativity, 26, 3, \dodoi{10.1007/s41114-023-00044-3}

\bibitem[{Rossetto {et~al.}(2023)Rossetto, Frauendiener, Brunet, \& Melatos}]{rossetto2023Magneticallya}
Rossetto, P. H.~B., Frauendiener, J., Brunet, R., \& Melatos, A. 2023, Monthly Notices of the Royal Astronomical Society, 526, 2058, \dodoi{10.1093/mnras/stad2850}

\bibitem[{Singh {et~al.}(2020)Singh, Haskell, Mukherjee, \& Bulik}]{singh2020Asymmetric}
Singh, N., Haskell, B., Mukherjee, D., \& Bulik, T. 2020, Monthly Notices of the Royal Astronomical Society, 493, 3866, \dodoi{10.1093/mnras/staa442}

\bibitem[{Sur \& Haskell(2021)}]{sur2021Impact}
Sur, A., \& Haskell, B. 2021, Publications of the Astronomical Society of Australia, 38, e043, \dodoi{10.1017/pasa.2021.39}

\bibitem[{Suvorov \& Melatos(2019)}]{suvorov2019Relaxation}
Suvorov, A.~G., \& Melatos, A. 2019, Monthly Notices of the Royal Astronomical Society, 484, 1079, \dodoi{10.1093/mnras/sty3518}

\bibitem[{Suvorov \& Melatos(2020)}]{suvorov2020Recycled}
---. 2020, Monthly Notices of the Royal Astronomical Society, 499, 3243, \dodoi{10.1093/mnras/staa3132}

\bibitem[{Thorne(1980)}]{thorne1980Multipole}
Thorne, K.~S. 1980, Reviews of Modern Physics, 52, 299, \dodoi{10.1103/RevModPhys.52.299}

\bibitem[{Ushomirsky {et~al.}(2000)Ushomirsky, Cutler, \& Bildsten}]{ushomirsky2000Deformations}
Ushomirsky, G., Cutler, C., \& Bildsten, L. 2000, Monthly Notices of the Royal Astronomical Society, 319, 902, \dodoi{10.1046/j.1365-8711.2000.03938.x}

\bibitem[{Vigelius \& Melatos(2008)}]{vigelius2008Threedimensional}
Vigelius, M., \& Melatos, A. 2008, Monthly Notices of the Royal Astronomical Society, 386, 1294, \dodoi{10.1111/j.1365-2966.2008.13139.x}

\bibitem[{Vigelius \& Melatos(2009)}]{vigelius2009Resistive}
---. 2009, Monthly Notices of the Royal Astronomical Society, 395, 1985, \dodoi{10.1111/j.1365-2966.2009.14698.x}

\bibitem[{Wette {et~al.}(2010)Wette, Vigelius, \& Melatos}]{wette2010Sinking}
Wette, K., Vigelius, M., \& Melatos, A. 2010, Monthly Notices of the Royal Astronomical Society, 402, 1099, \dodoi{10.1111/j.1365-2966.2009.15937.x}

\end{thebibliography}
